\documentclass[a4paper, 11pt, oneside]{article}
\pdfoutput=1
\usepackage{jheppub}		
\usepackage{graphicx,wrapfig,float,slashed,indentfirst}
\usepackage{amsmath}
\allowdisplaybreaks
\usepackage{xcolor,bbm}
\usepackage{fancyhdr}	
\usepackage{epstopdf}
\usepackage{afterpage}
\usepackage{xr}
\usepackage{graphics}
\usepackage[mathscr]{eucal}
\usepackage{xspace}
\usepackage{listings}
\usepackage{gensymb}
\usepackage{pstricks}
\usepackage{soul}
\usepackage{xspace}
\usepackage{braket}
\usepackage{slashed}
\usepackage{subfigure}
\usepackage{array}
\usepackage[force]{feynmp-auto}
\usepackage[makeroom]{cancel}
\newcommand{\tetab}{\renewcommand{\arraystretch}{1.3}}

\newcommand{\nc}{\newcommand}

\nc{\beq}{\begin{equation}}  \nc{\eeq}{\end{equation}}
\nc{\bea}{\begin{eqnarray}}  \nc{\eea}{\end{eqnarray}}
\nc{\baa}{\begin{array}}     \nc{\eaa}{\end{array}}
\nc{\bit}{\begin{itemize}}   \nc{\eit}{\end{itemize}}
\nc{\ben}{\begin{enumerate}} \nc{\een}{\end{enumerate}}
\nc{\bce}{\begin{center}}    \nc{\ece}{\end{center}}
\nc{\bpm}{\begin{pmatrix}}   \nc{\epm}{\end{pmatrix}}
\nc{\bvt}{\begin{verbatim}}  \nc{\evt}{\end{verbatim}}
\nc{\non}{\nonumber} 
\newcolumntype{M}{>{$\vcenter\bgroup\hbox\bgroup}c<{\egroup\egroup$}}

\def\gev{\;\hbox{GeV}}

\def\diag{\hbox{\diag}}
\def\zBB{{\mathbb{Z}}}

\def\z2{\zBB_2}

\def\mone{m_1}
\def\mtwo{m_2}
\def\zp{X}
\def\mzp{M_{X}}
\def\mx{M_{X}}
\def\vx{v_x}
\def\gx{g_x}

\def\uone{U(1)_X}

\def\vr{v}
\def\sigv{\langle\sigma\vr\rangle}

\def\vr{v_{\rm rel}}
\def\pthat{\widehat}

\def\lsim{\mathrel{\raise.3ex\hbox{$<$\kern-.75em\lower1ex\hbox{$\sim$}}}}
\def\gsim{\mathrel{\raise.3ex\hbox{$>$\kern-.75em\lower1ex\hbox{$\sim$}}}}

\def\ot#1{%
  \mathrel{\vbox{\offinterlineskip\ialign{%
    \hfil##\hfil\cr
    $\scriptscriptstyle(\,\sim\,)$\cr
    \noalign{\kern-.1ex}
    $#1$\cr
}}}}

\def\inv#1{\frac1{#1}}
\newcommand{\marrow}[5]{%
    \fmfcmd{style_def marrow#1
    expr p = drawarrow subpath (1/4, 3/4) of p shifted 6 #2 withpen pencircle scaled 0.4;
    label.#3(btex #4 etex, point 0.5 of p shifted 6 #2);
    enddef;}
    \fmf{marrow#1,tension=0}{#5}}

\begin{document}

\begin{flushright}
MAN/HEP/2018/004\\
December 2018\\[-2cm]
${}$
\end{flushright}

\title{Gauge-Independent Approach to Resonant Dark Matter Annihilation}
\author[1]{Mateusz Duch}
\author[1]{\hspace{-1.5mm}, Bohdan Grzadkowski}
\author[2]{and Apostolos Pilaftsis}
\affiliation[1]{Faculty of Physics, University of Warsaw, Pasteura 5, 02-093 Warsaw, Poland}
\affiliation[2]{Consortium for Fundamental Physics, School of Physics and Astronomy,\\
Oxford Road, University of Manchester, Manchester M13 9PL, U.K.}
\emailAdd{mateusz.duch@fuw.edu.pl}
\emailAdd{bohdan.grzadkowski@fuw.edu.pl}
\emailAdd{Apostolos.Pilaftsis@manchester.ac.uk}
\date{\today}

\abstract{In spontaneously broken gauge theories, transition
  amplitudes describing dark-matter (DM) annihilation processes
  through a resonance may become highly inaccurate close to a
  production threshold, if~a~Breit--Wigner (BW) ansatz with a constant
  width is used. To partially overcome this problem, the BW propagator
  needs to be modified by including a momentum dependent decay width.
  However, such an approach to resonant transition amplitudes
  generically suffers from gauge artefacts that may also give rise to
  a bad or ambiguous high-energy behaviour for such amplitudes. We
  address the two problems of gauge dependence\- and high-energy
  unitarity within a gauge-independent framework of resummation
  implemented by the so-called Pinch Technique. We study DM
  annihilation via scalar resonances in a gauged U(1)$_X$
  complex-scalar extension of the Standard Model that features a
  massive stable gauge field which can play the role of the DM. We
  find that the predictions for the DM abundance may vary
  significantly from previous studies based on the naive BW ansatz 
  and propose an alternative simple approximation which leads to the 
  correct DM phenomenology. In particular, our results do not depend 
  on the gauge-fixing parameter and are consistent with considerations 
  from high-energy unitarity. }


\maketitle

\section{Introduction}

Unveiling the nature of Dark Matter~(DM) constitutes one of the
biggest challenges in Cosmology and possibly in Particle
Physics. Even though several pieces of evidence coming from both
cosmological and astrophysical scales confirm its presence, the actual
composition of the DM itself remains elusive to us thus far. In
numerous models, the DM is assumed to be a new kind of
massive particles that were in thermal\- equilibrium with the Standard
Model~(SM) particles in the early Universe. In such thermal\- DM
models, the DM relic abundance crucially depends on the rate at which
DM particles annihilate into thermal bath states. This rate is
proportional to the thermally averaged DM-pair-annihilation
cross-section summed over all possible final states. Most
remarkably${}$, if this cross-section has a value which is typical to
the one governing SM weak interactions, then the predicted DM density
turns out to be in the right ballpark in agreement with
observations. Note that the most accurate determination of the DM
relic density comes from measurements of the Cosmic Microwave
Background (CMB) by the Planck satellite in conjunction with a number
of other experiments and astrophysical data (for a recent analysis,
see~\cite{Aghanim:2018eyx}).
  
A popular DM
scenario~\cite{Bento:2001yk,Deppisch:2008bp,MarchRussell:2008tu,Ibe:2008ye,
  Ibe:2009dx,Ibe:2009en,
  Guo:2009aj,Bi:2009uj,Backovic:2009rw,Braaten:2013tza,Campbell:2015fra,
  Choi:2017mkk,Duch:2017nbe,Bai:2017fav,Chu:2018fzy} is based on the
working hypothesis that there exists a mediator with couplings to both
DM and SM particles, and whose mass happens to be approximately twice
that of the~DM. In such scenarios, one has to consider DM annihilation
processes at energies where the tree-level propagator of the mediator
becomes singular.  A~standard way to avoid such singularities is to
employ a Breit--Wigner (BW) ansatz for the propagator with a constant
decay width~\cite{Breit:1936zzb}.  Then, the amplitude gets
regularized by the non-zero width of the mediator and can thus get
drastically enhanced by many orders of magnitude, when compared to
other non-resonant contributions. In Quantum Field Theory (QFT), the
BW form of the propagator usually arises from a Dyson series summation
of self-energy graphs of the mediator. In the so-called on-mass-shell
(OS) scheme of renormalization~\cite{Sirlin:1980nh}, the dispersive
parts of the self-energies renormalize the masses, whilst their
absorptive part is related to the decay width of the mediator.

From the phenomenological point of view, the channel of resonant DM
annihilation turns out to be an attractive option, as it leads to
suppressed DM-nucleon cross-sections thereby avoiding the tight
constraints emanating from the null results of direct DM
searches. Therefore, it should not be too surprising that this
resonant region in question\- may become the only viable region in the
parameter space of a given model with DM mass in the GeV range that
survives after all direct detection limits on the DM-nucleon
cross-section were imposed \cite{Athron:2018hpc,Balazs:2017ple}.

As was first noted in~\cite{Duch:2017nbe}, the BW approximation with a
constant decay width in the propagator of the mediator can become very
inaccurate close to the DM production threshold, especially when the
respective DM channel contributes significantly to the decay width of
the mediator. To partially remedy this problem, one is compelled to
use an effective BW propagator with momentum-dependent or running
width for the exchanged particle in the $s$-channel. In this article,
we will go beyond the previously used non-relativistic
approach~\cite{Duch:2017nbe}. In general, an $s$-dependent width
results from the imaginary (absorptive) part of the self-energy of the
mediator. This quantity is only gauge-independent when evaluated at
the pole of the propagator, but becomes gauge-variant in the off-shell
region. This was a well-known problem in QFT and pertains to the
question whether a consistent gauge-independent definition of
off-shell Green's function for unstable particles exists in
spontaneously broken gauge theories~\cite{Pilaftsis:1989zt}. To deal
with this issue, a number of recipes and methods have been put forward
by several authors, such as the Laurent series
expansion~\cite{Stuart:1991xk,Sirlin:1991fd}, the complex mass
scheme~\cite{Nowakowski:1993iu, Denner:2006ic}, the fermion loop
scheme~\cite{Argyres:1995ym}, and the effective theory
approach~\cite{Beneke:2003xh}.

An elegant and equally consistent gauge-independent framework to
address the afore-mentioned problem is the so-called Pinch
Technique~(PT)~\cite{Cornwall:1981zr,Cornwall:1989gv,Papavassiliou:1995fq,
  Papavassiliou:1995gs,Binosi:2009qm}. The PT preserves basic
properties of QFT, such as analyticity, unitarity and the gauge
invariance of the classical action. The PT resummation approach to
unstable particles~\cite{Papavassiliou:1995fq,Papavassiliou:1995gs,Pilaftsis:1997dr}
was extensively studied in the literature originally within the
context of the SM~\cite{Papavassiliou:1989zd,Degrassi:1992ue,Degrassi:1993kn,
  Papavassiliou:1997fn,Papavassiliou:1997pb} and more recently in two
Higgs-doublet models~\cite{Ellis:2004fs,Krause:2016oke,Krause:2017mal,Chiang:2017vvo}.

In this paper we discuss the problems that arise in the relativistic
treatment of resonant DM annihilation processes in spontaneously
broken gauge theories and show how these can be avoided in a
resummation approach implemented by the PT. As an archetypal\- model,
we consider a gauged U(1)$_X$ complex-scalar extension of the SM that
includes a massive stable gauge boson $X$ as a candidate particle for
a Vector DM (VDM).  We explicitly demonstrate how the transition
amplitude for DM-pair annihilation has the proper high-energy
unitarity limit in compliance with the Equivalence
Theorem~\cite{Cornwall:1974km, Lee:1977eg} and its generalized
version~(GET)~\cite{Chanowitz:1985hj,Gounaris:1986cr}.  By virtue of
the PT resummation method adopted in this paper, we obtain predictions
for the yield of DM abundance that may vary significantly from
previous${}$ considerations based on the naive BW approximation. In
particular, we illustrate the gauge independence of our results and
their consistency with high-energy unitarity. In the same context, we
present an alternative simple approximation which leads to the correct
DM phenomenology.

The paper is organized as follows. In sec.~\ref{resonances} we discuss
the process of resonant DM-pair annihilation and the problems that
arise in the relativistic treatment of a VDM in the vicinity of a
resonance. In sec.~\ref{resummation} we describe the PT resummation
approach and apply it to transition amplitudes that are typical to DM
annihilation processes, such as $XX \to \bar{f} f$, where~$f$ is a SM
fermion. By means of this approach, we obtain a Born-improved
amplitude which is gauge independent and possesses the proper
asymptotics in the high-energy limit. In sec.~\ref{reldens} we use the
Born-improved amplitude derived in the previous section to
compute the annihilation cross-sections and DM relic density.
This enables us to assess the 
significance of our results by comparing them with those derived with
other methods. In sec.~\ref{con} the key points of our analysis were
summarised. Technical details of this study were relegated to a number
of appendices. Specifically, appendix~\ref{model} provides further
details of the VDM model under study, appendix~\ref{bfgfeyn} presents
the Feynman rules for this model in the $R_\xi$ and background
field gauges (BFG), and appendix~\ref{ptvertices} contains analytical
expressions of one-loop vertex corrections in the PT. Finally,
appendix~\ref{getapp} complements our proof of the GET for the DM
annihilation process $XX \to \bar{f} f$.

\section{Resonant Dark Matter Annihilation}
\label{resonances}

Annihilation of DM in the vicinity of a resonance has certain features
that renders it distinct from a generic scenario of a thermal DM. First,
the thermally averaged cross-section displays${}$ strong dependence on the
temperature~$T$ that can lead to a prolonged period of effective DM
annihilation which substantially changes the comoving DM density, even
though the DM itself may have already decoupled from the thermal
bath. Second, the cross-section may be significantly enhanced at small
velocities leading to a strengthening of the signal probed by DM
indirect searches~\cite{Ibe:2008ye,Guo:2009aj}. Finally, the coupling
of the DM to SM particles can become rather suppressed. This last
property can be quite challenging for collider or direct detection
experiments as it usually gives rise to weaker constraints for this
area of parameter space. It~also raises the temperature of kinetic
decoupling which, in connection with the high $T$-variability of the
annihilation cross-section, may affect the predictions of relic
density~\cite{Bi:2011qm,Duch:2017nbe,Binder:2017rgn}.

In the Born approximation, the transition amplitude for a resonant DM
annihilation process does not describe consistently the dynamics of a possible
unstable mediator with mass~$M$, because\- its tree-level propagator
$\Delta^0(s) \equiv (s-M^2)^{-1}$ becomes singular at~$s = M^2$.  The
standard way to treat this singularity is to perform a Dyson series
summation of the mediator's self-energy~${\Pi (s)}$ which results in
a replacement of the tree-level propagator~$\Delta^0(s)$ with a
resummed one, $\Delta (s) \equiv (s-M^2+\Pi(s))^{-1}$.  In the OS
scheme of renormalization~\cite{Sirlin:1980nh}, this modification
amounts to the well-known BW approximation~\cite{Breit:1936zzb}, with
${\Pi(s)\approx\Im m\,\Pi(s=M^2)=iZ^{-1}\,M\Gamma}$, where~$Z$ is the
wave-function renormalization (set to~1 at the one-loop level),
and~$M$ and~$\Gamma$ are the renormalized mass and width of the resonant
particle in this scheme. In this way, one obtains a finite analytic
expression for the amplitude in the resonance region.  For instance,
the cross-section for a $2\to 2$ annihilation process, which proceeds
via the $s$-channel and has two identical DM particles of mass $m_i$
in the initial state and two particles of equal mass $m_f$ in the
final state, is approximately given by
\begin{equation}
   \label{csann}
\sigma \ \simeq\  \frac{1}{s}\sum_{f\neq i}\frac{M^2 \Gamma^2 B_i
  B_f}{(s-M^2)^2\: +\: M^2 \Gamma^2}\; .
\end{equation}
In the above, we followed the notation and conventions of
\cite{Ibe:2008ye}, after taking into account an extra factor of~2
because of identical particles in the initial state. Hence, $B_i$ and
$B_f$ in~\eqref{csann} are the branching ratios for the mediator state
to decay into initial and final states, denoted as $i$ and~$f$, respectively.

Nevertheless, when the dominant contribution to the mediator's
self-energy~$\Pi (s)$ comes from particles which are also the initial
states of the transition amplitude, then the BW approximation is in
general not applicable~\cite{Duch:2017nbe}. In this case, one finds
high inaccuracies in the transition amplitudes that result from the
phase space factor $\sqrt{1-s/(4m^2)}$ present in $\Pi(s)$, which
varies substantially for $s\stackrel{>}{{}_\sim} 4m^2$.  To deal with this problem,
one has to consider the energy-dependence of the imaginary part of the
self-energy, which is sometimes described by the running width of the
mediator $\Gamma(s)\equiv\Im m\,\Pi(s)/M$.  However, this approach
encounters other serious theoretical problems. As the resummation
process relies on taking a subset of graphs from each order of the
perturbative expansion, the subtle cancellations which guarantee the
gauge-independence of the amplitudes at each order of the expansion
are spoiled and the resulting expression has explicit dependence on
the gauge fixing parameter(s) chosen to calculate $\Pi(s)$.  The
gauge-dependent amplitude results in an ambiguous annihilation
cross-section, and as such its physical relevance becomes rather obscured.

In order to address the above issue, we adopt the PT framework which enables
us to calculate such resonant DM annihilation amplitudes following a
gauge-independent and self-consistent approach. As mentioned in the
introduction, the so-derived Born-improved amplitudes will satisfy
several desirable field-theoretic properties as a consequence of the
analyticity, unitarity and gauge-invariance of the $S$-matrix. Our
approach implemented by the PT will be illustrated within a VDM
model~\cite{Hambye:2008bq,Lebedev:2011iq,Farzan:2012hh,Baek:2012se,Baek:2014jga,
  Duch:2015jta}. This is an extension of the SM augmented by local
U(1)$_X$ symmetry and by a complex scalar field~$S$.  The field~$S$
has a portal interaction with the SM Higgs doublet, and also develops
a non-zero vacuum expectation value (VEV) that breaks~U(1)$_X$
spontaneously. The resulting gauge boson~$X$ acquires a mass due to the
Higgs mechanism and can be made stable, provided its kinetic mixing
with the hypercharge gauge boson is forbidden by a~$Z_2$ symmetry.  In
this~U(1)$_X$ extension of the SM, the vector boson~$X$ can play the
role of a viable DM.  This VDM model contains two scalar mass
eigenstates, denoted as $h_1$ and $h_2$, where the first scalar
represents the SM-like Higgs boson, while the second one will serve as
a resonant mediator. The couplings of the $h_1$ and $h_2$ scalars to
SM matter and VDM are governed by the interaction Lagrangian,
\begin{equation}
\begin{split}
\mathcal{L}_{\rm int}\ = &\ \frac{h_1 \cos \alpha + h_2 \sin
  \alpha}{v}\,\Big( 2
  M_W^2 W_\mu^+ W^{\mu-} + M^2_Z Z_\mu Z^\mu - \sum_f m_f \bar{f}f\Big)\\ 
&+\: 2 g_x \mzp\,\Big(\!-h_1 \sin \alpha + h_2 \cos \alpha\Big)\, X_\mu X^\mu\,,
\end{split}
\end{equation}
where $\alpha$ is the scalar mixing angle and $\gx$ is the U(1)$_X$
gauge coupling.  Our primary focus in this analysis will be on the
resonance region, where $m_2 \approx 2\mx$ with a scalar mixing angle
$\alpha\ll 1$. Further details of the VDM model under consideration may be
found in appendix~\ref{model}.

\section{Gauge Independence and High Energy Unitarity}
\label{resummation}

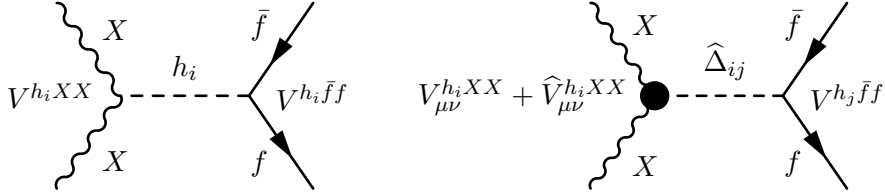
\begin{figure}[t!]
\begin{center}
\begin{tabular}{cc}	        
\begin{fmffile}{xxffTR}
	\begin{fmfgraph*}(120,70)
	   	\fmfleft{i,j}
        \fmfright{k,l}
      	\fmf{boson,label=$\zp$}{i,v1}
	   	\fmf{boson,label=$\zp$}{j,v1}
	   	\fmfv{l.d=10,label=$V^{h_iXX}$}{v1}
	   	\fmf{dashes,label=$h_i$}{v2,v1}
	   	\fmfv{label=$V^{h_i\bar ff}$,l.d=10}{v2}
        \fmf{fermion,label=$f$,l.s=right}{v2,k}
	   	\fmf{fermion,label=$\bar{f}$,l.s=right}{l,v2}
	\end{fmfgraph*}
\end{fmffile}
& \hspace{2cm}\begin{fmffile}{xxffBI}
	\begin{fmfgraph*}(120,70)
	   	\fmfleft{i,j}
        \fmfright{k,l}
      	\fmf{boson,label=$\zp$}{i,v1}
	   	\fmf{boson,label=$\zp$}{j,v1}
	   	\fmfv{decor.shape=circle,decor.filled=full,decor.size=5thick,
	   		l.d=10,
	   		label=$V^{h_iXX}_{\mu\nu}+\pthat V^{h_iXX}_{\mu\nu}$}{v1}
	   	\fmfv{label=$V^{h_j\bar ff}$,l.d=10}{v2}
	   	\fmf{dashes,label=$\pthat\Delta_{ij}$}{v2,v1}
       \fmf{fermion,label=$f$,l.s=right}{v2,k}
	   	\fmf{fermion,label=$\bar{f}$,l.s=right}{l,v2}
	\end{fmfgraph*}
\end{fmffile}
\end{tabular}  
\end{center}
\caption{\label{feyngraphsxxff}The tree-level (left) and the
  Born-improved (right) Feynman diagram for the process $XX\rightarrow
  f\bar f$.} 
\end{figure}

In this section, we will describe a gauge-invariant resummation
approach to resonant transition amplitudes, within the Pinch Technique
(PT) framework. As an illustrative example, we will be studying the
annihilation process $XX\rightarrow \bar ff$, where $X$ is a U(1)$_X$
gauge field which assumes the role of the DM and $f$ collectively
stands for a SM fermion.  In the U(1)$_X$ scenario under study, the DM
annihilation process, $XX\rightarrow \bar ff$, proceeds via the
exchange of a coupled system of two mixed scalar resonances $h_1$ and
$h_2$. As illustrated in the left diagram of fig. \ref{feyngraphsxxff},
the tree-level amplitude of such a reaction reads
\begin{equation} 
   \label{eq:Atree}
\epsilon^\mu(p_1) \epsilon^\nu(p_2)\, i\mathcal{A}^{XX\rightarrow \bar f f}_{\mu\nu } \
=\ \epsilon^\mu(p_1) \epsilon^\nu(p_2)\, V^{XXh_i}_{\mu\nu}\frac{i\delta_{ij}}{s-m^2_i}\; V^{h_j
  \bar f f}\,\bar{u}(r_2)v(r_1)\,.  
\end{equation}
Here, summation over the repeated indices $i,j = 1,2$ is implied,
$m_{1,2}$ are the masses of the Higgs scalars $h_{1,2}$, and
$V^{XXh_i}_{\mu\nu}$ and $ V^{h_j \bar f f}$ denote the tree-level
expressions for the $XXh_i$ and $h_j \bar f f$ vertices,
respectively. Moreover, $\epsilon^\mu(p_1)$ and $\epsilon^\nu(p_2)$
denote the polarizations of the $X$-vector bosons with four-momenta
$p_1$ and $p_2$, and $u(r_2)$ and $v(r_1)$ are the usual Dirac spinors
for the fermion-antifermion pair $f$ and $\bar{f}$ with momenta $r_2$
and $r_1$, respectively.

As stated in sec.~\ref{resonances}, the tree-level
amplitude~\eqref{eq:Atree} becomes singular near the resonance regions
$s \approx m^2_{1,2}$, requiring resummation of an infinite series of
self-energies~$\Pi_{ij}(s)$.  Our aim is to calculate a Born-improved
amplitude for the DM annihilation process $XX\rightarrow \bar ff$,
which does not depend on the gauge-fixing parameter $\xi_X$ used to
parameterize the unphysical degrees of freedom of the gauge field
$X$. In this context, we will adopt a gauge-independent resummation
approach (illustrated in the right diagram of fig.~\ref{feyngraphsxxff})
which is implemented by the PT in order to properly treat the coupled
system of the two resonances, $h_1$ and $h_2$, in a fashion analogous
to~\cite{Pilaftsis:1997dr}.  In this framework, in
sec.~\ref{sec:PTresum}, we obtain a resummed propagator which is a
$2\times 2$ matrix and is denoted as
$\pthat\Delta_{ij}$. Subsequently, in sec.~\ref{sec:GET} we discuss
the necessity of including one-loop corrections to the $XXh_i$
vertices and present the tree-level Ward Identities (WIs) satisfied by
the PT self-energies and PT vertices which ensure the validity of the
ET~\cite{Cornwall:1974km, Lee:1977eg}, as well as of the
GET~\cite{Chanowitz:1985hj,Gounaris:1986cr}.  Finally, we show in
sec.~\ref{sec:Asym} that the so-derived Born-improved amplitude
$XX\to \bar ff$ exhibits proper high-energy asymptotic behaviour as
expected from unitarity considerations.

\subsection{Pinch Technique Resummation}
\label{sec:PTresum}

  To start with, we first calculate the $X$-field contribution to the
  $h_ih_j$ self-energies~$\Pi_{ij}(s)$ in the VDM model of interest
  to us.   In the  renormalizable class of  $R_\xi$ gauges, 
upon inclusion of the would-be Goldstone bosons $G_X$ and the
pertinent ghost fields, 
 these are
  given by
\begin{eqnarray} 
   \label{pixi}
\Pi^{(XX)}_{ij}(s) &=& \frac{g_x
  ^2 R_{2i} R_{2j}}{32\pi^2\mx^2}\Big[\left( s^2-4\mx^2 s + 12 \mx^4
\right) B_0(s,\mx^2,\mx^2)\nonumber\\
&&\hspace{2cm} -\: (s^2-m^2_i m^2_j)
B_0(s,\xi_X\mx^2,\xi_X\mx^2)\Big]\,,
\end{eqnarray}
where $\xi_X$ is the gauge-fixing parameter associated with the
U(1)$_X$ gauge field. Resummation of this self-energy inevitably leads
to a gauge-dependent resonant amplitude and the appearance of
unphysical thresholds at $s = 4\,\xi_X M^2_X$, when $\xi_X \neq 1$.

We may now calculate $X$, $Z$, $W$, $t$, $h_k$, $h_l$ contributions to
the $h_ih_j$ self-energies~$\pthat{\Pi}_{ij}(s)$ employing
the PT, as done in~\cite{Papavassiliou:1997fn,Papavassiliou:1997pb}. For the case
of the VDM, these are given by
\begin{align} 
  \label{hatpi}
&\pthat\Pi^{(XX)}_{ij}(s) \ =\
\frac{g_x^2 R_{2i} R_{2j}}{8\pi^2}\left[\frac{(m_im_j)^2}{4
    \mzp^2}+\frac{m^2_i+m^2_j}{2}-(2s - 3\mzp^2) \right]
B_0(s,\mzp^2,\mzp^2)\;,\\
&\pthat\Pi^{(WW)}_{ij}(s) \ =\
\frac{g^2 R_{1i} R_{1j}}{16\pi^2}\left[\frac{(m_im_j)^2}{4
    \mzp^2}+\frac{m^2_i+m^2_j}{2}-(2s - 3M_W^2) \right]
B_0(s,M_W^2,M_W^2)\;,\\
&\pthat\Pi^{(ZZ)}_{ij}(s) \ =\
\frac{g^2 R_{1i} R_{1j}M_Z^2}{32\pi^2M_W^2}\left[\frac{(m_im_j)^2}{4
    \mzp^2}+\frac{m^2_i+m^2_j}{2}-(2s - 3M_Z^2) \right]
B_0(s,M_Z^2,M_Z^2)\;,\\
&\pthat\Pi^{(tt)}_{ij}(s) \ =\
\frac{3g^2 R_{1i} R_{1j}m_t^2}{32\pi^2M_W^2}\left(s - 4m_t^2\right)
B_0(s,m_t^2,m_t^2)\;,\\
&\pthat\Pi^{(h_k h_l)}_{ij}(s) \ =\
-\,\frac{V^h_{ikl}V^h_{jkl}}{32\pi^2}\,
B_0(s,m_{h_k}^2,m_{h_l}^2)\;,
\end{align}
where
\begin{equation}
   \label{eq:VtH}
B_0 (p^2,m^2_a,m^2_b)\ \equiv\ 
(2\pi \mu)^{4-n}\, \int\frac{d^nk}{i\pi^2}\, \frac{1}{(k^2-m^2_a)\,
  [(k+p)^2 - m^2_b]}\; 
\end{equation} 
is the 't Hooft--Veltman function~\cite{tHooft:1978jhc} defined in
$n = 4 -2\epsilon$ dimensions and the coupling $V^h_{ikl}$ is
specified in table~\ref{vertices}.  Note that the above results for
the PT self-energies~$\pthat{\Pi}_{ij}(s)$ are also in agreement with
that obtained in~\cite{Ellis:2004fs}, upon appropriately replacing the
couplings of that model. Moreover, the same results can be derived by
making use of the established equivalence between the PT and the
covariant Background Field Gauge~(BFG) for
$\xi_Q=1$~\cite{Denner:1994nn,Hashimoto:1994ct,
  Papavassiliou:1994yi,Pilaftsis:1996fh}. For reader's convenience,
the Feynman rules in the BFG are presented in
Appendix~\ref{bfgfeyn}. Evidently${}$, the $h_ih_j$
self-energies~$\pthat{\Pi}_{ij}(s)$ do not depend on the gauge-fixing
parameters $\xi_{X}$, $\xi_{W}$, $\xi_{Z}$ and all display an
absorptive part at the {\em physical} thresholds
$s = 4 M^2_X,\ 4 M^2_W,\ 4M^2_Z$, as expected.

The full PT self-energies $\pthat{\Pi}_{ij}(s)$ can then be
systematically resummed, yielding the PT resummed propagator
\begin{equation}
  \label{eq:Dhat}
i\pthat\Delta\ =\ i\Delta^0\: +\: i\Delta^0\,i\pthat\Pi\, i\Delta^0\:  +\:
i\Delta^0\,(i\pthat\Pi\, i\Delta^0)^2\: +\: \ldots  
\end{equation}
In the above, we have suppressed the indices $i,j = 1,2$ labelling the
Higgs scalars $h_{1,2}$ and the $s$-dependence of the propagators and
self-energies. Moreover,
$\Delta^0 (s) = {\rm diag}\big[ (s-m^2_1)^{-1},$ $(s-m^2_2)^{-1}\big]$
is the diagonal tree-level propagator matrix. From~\eqref{eq:Dhat},
one then gets
\begin{equation}
   \label{resprop}
\pthat \Delta (s)= \frac{1}{D(s)}\left(\begin{array}{cc}
 s-m_2^2 + \pthat\Pi_{22}(s) & -\pthat\Pi_{12} (s) \\ 
-\pthat\Pi_{21}(s)  & s-m_1^2 + \pthat\Pi_{11}(s)
\end{array}\right),
\end{equation}
with $D(s) = \big[s-m_1^2 + \pthat\Pi_{11}(s)\big]\,\big[(s-m_2^2 + \pthat\Pi_{22}(s)\big]\: -\:
\pthat\Pi_{12}(s)\pthat\Pi_{21}(s)$.
As we will see in the next section, in addition to its gauge independence,
a Born-improved amplitude for $XX\rightarrow \bar ff$ must have the
proper high-energy asymptotic behaviour, as dictated by the ET.

\subsection{The Generalized Equivalence Theorem}
\label{sec:GET}

Replacing naively the tree-level propagator matrix $\Delta^0 (s)$ with
the resummed one $\pthat \Delta (s)$ in~\eqref{eq:Atree} modifies the
transition amplitude~$\mathcal{A}^{XX\rightarrow \bar f f}(s)$ not
only in the vicinity of the resonance region, but also changes
drastically its high-energy limit as $s/(4M^2_X) \to \infty$. This is
a generic problem for most forms of BW
propagators~\cite{Valencia:1990jp,Seymour:1995np} and is due to the
presence of non-zero self-energies $\pthat \Pi_{ij}(s)$ or decay
widths in $\pthat \Delta (s)$ that may posses non-trivial
$s$-dependence. The latter distort subtle cancellations which are
triggered by the WIs that result from the gauge invariance of the
classical action, thus leading to a different high-energy limit from
the one expected in the Born approximation.

To ensure the proper high-energy asymptotic behaviour of a scattering
process involving massive gauge fields in the initial or final state,
the amplitude of such process must obey the
GET~\cite{Chanowitz:1985hj,Gounaris:1986cr}, which is a consequence of
the classical WIs mentioned above. Applying the GET to the amplitude
of the process $XX\rightarrow \bar f f$ gives
\begin{equation}
   \label{get}
\begin{split}
\mathcal{A}_{X_L(p_1)X_L(p_2)\rightarrow \bar f f}\ =\
&-\mathcal{A}_{G_X(p_1)G_X(p_2)\rightarrow \bar f f}\: -\: i\mathcal{A}_{x^\mu(p_1)G_X(p_2)\rightarrow \bar f f}\\
&-i\mathcal{A}_{G_X(p_1)x^\nu(p_2)\rightarrow \bar f
  f}\: +\: \mathcal{A}_{x^\mu(p_1)x^\nu(p_2)\rightarrow \bar f f}\; .
\end{split}
\end{equation}
Equation~\eqref{get} establishes a relation between the amplitude with
initial longitudinally polarized $X$-bosons, denoted here as
$X^\mu_L(p_1)=\epsilon^\mu_L(p_1)$ and
$X^{\nu}_L(p_2)=\epsilon^{\nu}_L(p_2)$, and the amplitudes with
corresponding would-be Goldstone bosons $G_X(p_{1,2})$ or the
energetically suppressed remainders, $x^\mu(p_1)$ and $x^{\nu}(p_2)$.
Specifically,  $x^\mu(p_1)$  is defined as
\begin{equation}
x^\mu(p_1)\ \equiv\ \epsilon^\mu_L(p_1)\: -\: \frac{p^\mu_1}{M_X}\; ,
\end{equation}
which satisfies the identities
\begin{equation}
   \label{xident}
x_\mu(p_{1})\,p_1^\mu\ =\ -M_X\,,\qquad x_\mu(p_1)\,x^\mu(p_1)\ =\ 0\; .
\end{equation}
Note that an analogous definition and set of identities hold for
$x^\nu(p_2)$ as well.

Proceeding as in~\cite{Papavassiliou:1997fn,Papavassiliou:1997pb}, we
may reinforce the validity of the GET  stated in~\eqref{get} by
considering one-loop $h_iXX$, $h_iX G_X$ and $h_i G_X G_X$
vertices within the PT framework.  More explicitly, the following
substitutions for the tree-level $h_i$-couplings need to be considered:
\begin{align}
  \label{mod_verhXX}
V^{h_iXX}_{\mu\nu}\ &\rightarrow\ \pthat \Gamma^{h_iXX}_{\mu\nu}(q,p_1,p_2)\: =\:
V^{h_iXX}_{\mu\nu}\: +\: \pthat V^{h_iXX}_{\mu\nu}(q,p_1,p_2),\\
  \label{mod_verhXG}
V_\mu^{h_i\zp G_\zp}\ &\rightarrow\ \pthat
                       \Gamma^{h_iXX}_{\mu\nu}(q,p_1,p_2)\: =\:
                        V_\mu^{h_i\zp G_\zp} \: +\: \pthat
                       V_\mu^{h_i\zp G_\zp}(q,p_1,p_2),\\
  \label{mod_verhGG}
V^{h_i G_\zp G_\zp}\ &\rightarrow\  \pthat \Gamma^{h_i
                      G_\zp G_\zp}(q,p_1,p_2) \: =\: V^{h_i G_\zp
                       G_\zp}\: +\:  \pthat V^{h_i
                      G_\zp G_\zp}(q,p_1,p_2)\,, 
\end{align} 
where the relevant PT one-loop vertices, $\pthat V^{h_iXX}_{\mu\nu}$,
$\pthat V_\mu^{h_i\zp G_\zp}$ and $\pthat V^{h_i G_\zp G_\zp}$, were
calculated by means of the BFG method and presented in Appendix
\ref{ptvertices}. One can show that when PT resummed vertices 
as given in \eqref{mod_verhXX}--\eqref{mod_verhGG} are employed, the 
relation~\eqref{get} is satisfied in terms of the PT
resummed (Born-improved) amplitudes.  

The proof of~\eqref{get} for the Born-improved amplitudes relies on
the fact that the PT one-loop vertices
$\pthat V^{h_i\zp\zp}_{\mu\nu}$, $\pthat V_\mu^{h_i\zp G_\zp}$ and
$\pthat V^{h_i G_\zp G_\zp}$, as well as the PT self-energies $\pthat \Pi_{ij}$,
$\pthat \Pi^{X G_X}_\mu$ and $\pthat \Pi^{G_X G_X}$, satisfy tree-like WIs
that are identical to those derived from the classical action. In
detail, these tree-like PT WIs read
\begin{align}
 \label{WI1}
&p^\nu_2 \pthat V^{h_i\zp\zp}_{\mu\nu} (q,p_1,p_2) + i \mzp \pthat
  V_\mu^{h_i\zp G_\zp}(q,p_1,p_2)\ =\ -\gx  R_{2i} \pthat\Pi^{X
  G_X}_\mu(p_1)\,,\\[3mm]
 \label{WI2}
&p_1^\mu \pthat V_\mu^{h_i\zp G_\zp}(q,p_1,p_2) + i \mzp \pthat V^{h_i
  G_\zp G_\zp}(q,p_1,p_2) = -\gx\Big[ R_{2j}\pthat\Pi_{ji}(q^2)
  + R_{2i}\pthat \Pi^{G_X G_X}(p_2)\Big],\\[-2mm] 
 \label{WI3}
& p_1^\mu p_2^\nu \pthat V^{h_i\zp\zp}_{\mu\nu} (q,p_1,p_2) + \mzp^2
  \pthat V^{h_i G_\zp G_\zp}(q,p_1,p_2)\nonumber\\
& \hspace{3cm} =\ i \gx \mzp \Big[ R_{2j}\pthat\Pi_{ji}(q^2) +
  R_{2i}\Big(\pthat \Pi^{G_X G_X}(p_1)+\pthat \Pi^{G_X
  G_X}(p_2)\Big)\Big]\;,\\[4mm]
 \label{WI4}
& \pthat\Pi_\mu^{XG_X}(p)\ =\ -\frac{iM_Xp_\mu}{p^2}\pthat\Pi^{G_XG_X}(p^2)\,.
\end{align}
It is not difficult to verify that the PT WIs~\eqref{WI1}--\eqref{WI4}
are indeed satisfied by the tree-level couplings of the theory,
$V^{h_i\zp\zp}_{\mu\nu}$, $V_\mu^{h_i\zp G_\zp}$ and
$V^{h_i G_\zp G_\zp}$, after making the replacements:
\begin{displaymath}
\pthat\Pi_{ij}(s) \ \to\ \delta_{ij}(s-m_i^2)\,,\qquad
\pthat\Pi_\mu^{X G_X} (p)\ \to\ iM_Xp_{\mu}\,,\qquad
\pthat\Pi^{G_X G_X}(p)\ \to\ -p^2\;.
\end{displaymath}  
Further details pertinent to the proof of~\eqref{get} are given in Appendix~\ref{getapp}.

\subsection{High-Energy Asymptotics of the Born-Improved Amplitude}
\label{sec:Asym}

We will now study the high-energy behaviour of the PT resummed
amplitude for the process $XX \to \bar f f$. The high-energy
asymptotics of such an amplitude is similar to the one of
the SM process $\bar f f \rightarrow ZZ$ mediated by the Higgs boson,
which was considered in~\cite{Papavassiliou:1997fn,Papavassiliou:1997pb} and studied in
more detail in~\cite{Papavassiliou:1999qn}.

As we are interested in deriving a minimal Born-improved amplitude, we will ignore the
real (dispersive) parts of the self-energies and vertex
corrections. In the OS scheme, these are either subdominant or
suppressed by higher orders in the resonance region~\cite{Pilaftsis:1997dr}. Hence, we 
will only consider the imaginary (absorptive) parts of the PT resummed
propagators $\pthat \Delta_{ij} (s)$ and the PT vertices
$\pthat \Gamma^{h_iXX}_{\mu\nu}$. 
In this minimal self-consistent PT framework,  the Born-improved
amplitude takes on the form
\begin{equation}
   \label{amp}
   \begin{split}
i\mathcal{\pthat A}_{XX\rightarrow \bar f f}\ &\equiv\ 
\epsilon^\mu(p_1) \epsilon^\nu(p_2)\, i\mathcal{\pthat A}_{\mu\nu}^{XX\rightarrow \bar f f}\\
&=\ \epsilon^\mu(p_1) \epsilon^\nu(p_2)\;\pthat
\Gamma^{h_iXX}_{\mu\nu}(q,p_1,p_2)\, i\pthat\Delta_{ij}(s)\, V^{h_j \bar f
  f}\bar{u}(r_2)v(r_1)\; ,
  \end{split}
\end{equation}
where $\pthat \Gamma^{h_iXX}_{\mu\nu}(q,p_1,p_2) =
V^{h_iXX}_{\mu\nu} + \pthat V^{h_iXX}_{\mu\nu}(q,p_1,p_2)$ are the PT
one-loop vertices, whose absorptive parts 
are given in Appendix~\ref{ptvertices}. 

To analyse the high-energy asymptotics of
$\mathcal{\pthat A}_{XX\rightarrow \bar f f}$, we consider the
longitudinal $X$ boson polarizations which in the centre-of-mass (CoM)
frame can be expanded as follows
\begin{equation}
\epsilon_{L}^{\mu}(p_1) = \frac{p_1^\mu}{M_X} - 2M_X \frac{p_2^\mu}{s}\
+\ \mathcal{O}\left(\!\frac{16 M_X^4}{s^2}\!\right)\,,\qquad
\epsilon_{L}^{\mu}(p_2)=\frac{p_2^\mu}{M_X} - 2M_X\frac{p_1^\mu}{s}\ +\
\mathcal{O}\left(\!\frac{16 M_X^4}{s^2}\!\right)\; . 
\end{equation} 
For instance, taking into account the absorptive effects due to the opening
of the $XX$ threshold~only, the amplitude exhibits the following
asymptotic behaviour:
\begin{align}
   \label{asymp_ful}
\mathcal{\pthat A}^{(XX)}_{X_L(p_1)X_L(p_2) \rightarrow
  \bar f f}\ &=\ g\gx m_f \bar{u}(r_2)v(r_1)\sin(2\alpha)(m_1^2-m_2^2)\nonumber\\
&\times\, \frac{64  \pi \mx^2-i6\gx^2\big(m_1^2\sin^2\alpha +m_2^2
  \cos^2\alpha\big)}{128\pi M_W\mx^3 s}\ +\ \mathcal{O}\left(\!\frac{16 M_X^4}{s^2}\!\right)\,,
\end{align}
where $g$ is SU(2)$_L$ gauge coupling and $M_W$ is the $W^\pm$-boson mass.
This should be contrasted with the corresponding tree-level result,
\begin{equation}
   \label{asymp_tree}
\mathcal{A}^{\rm tree}_{X_L(p_1)X_L(p_2) \rightarrow
  \bar f f}\ =\ 
\frac{ig\gx m_f \bar{u}(r_2)v(r_1)
  \sin(2\alpha)(m_1^2-m_2^2)}{2sM_W\mx}\ +\
\mathcal{O}\left(\!\frac{16 M_X^4}{s^2}\!\right)\,. 
\end{equation}
It is not difficult to see that up to higher order terms
${\cal O}(g\gx^3)$, the PT resummed amplitude given
in~(\ref{asymp_ful}) approaches the tree-level result 
(\ref{asymp_tree}) in the high-energy limit. Evidently, the
Born-improved amplitude~(\ref{amp}) displays the expected
high-energy asymptotics.

It is instructive to see how the energetically constant terms
$\propto s^0$ cancel in the high-energy limit of both the tree-level
and Born-improved amplitude by virtue of the PT WIs. In the Born
approximation, such cancellation is a consequence of the orthogonality
of the mixing matrix $R$, since
\begin{equation}
   \label{eq:HEtree}
s\, R_{2i}\,\Delta^0_{ij}(s)\, R_{1j}\overset{s\rightarrow\infty}=\
R_{2i}\delta_{ij}R_{1j}\ =\ 0\;.
\end{equation}
Beyond the Born approximation, we may employ the PT WI~\eqref{WI3} to
derive an equivalent WI for the PT resummed vertex
$\widehat{\Gamma}^{h_i\zp\zp}_{\mu\nu}$ in the high-energy limit,
\begin{equation}
   \label{eq:WI3resum}
p_1^\mu p_2^\nu \pthat \Gamma^{h_i\zp\zp}_{\mu\nu} (q,p_1,p_2)\ =\ 
i \gx \mzp\, R_{2j}\pthat\Delta^{-1}_{ji}(q^2) \ +\ {\cal O}\big[\ln (s/M^2_X)\big]\;,
\end{equation}
where we have ignored contributions from
$\pthat V^{h_iG_XG_X}(q,p_1,p_2)$ and $\pthat\Pi^{G_XG_X}(p_{1,2})$
that grow as $\ln (s/M^2_X)$ or go to $s^0$, for
$p^2_1 = p^2_2 = M^2_X$~\cite{Papavassiliou:1999qn}. With the help
of~\eqref{eq:WI3resum}, the high-energy limit of the Born-improved
amplitude for the longitudinal $X$ bosons may then easily be evaluated
as follows:
\begin{equation}
\mathcal{\pthat A}_{X_L(p_1)X_L(p_2)\rightarrow \bar f f}\ \propto\
R_{2i}\, \pthat\Delta^{-1}_{ik}(s)\, \pthat\Delta_{kj}(s)\,R_{1j}
\overset{s\rightarrow\infty}=\ R_{2i}\delta_{ij}R_{1j}\ =\ 0\;.
\label{eq:limampff}
\end{equation}
Consequently, the leading high-energy asymptotics of
$\mathcal{\pthat A}_{X_L(p_1)X_L(p_2)\rightarrow \bar f f}$ is
expected to be proportional to $s^{-1}\ln (s/M^2_X)$ or $s^{-1}$, as
given in~\eqref{asymp_ful}. The energetically subleading terms are not
arbitrary either, but obey the GET stated in~\eqref{get}
[cf. Appendix~\ref{getapp}].

\begin{figure}[t!]
\begin{center}
\begin{tabular}{cc}	        
\hspace{1.8cm}\begin{fmffile}{xxvvBI}
	\begin{fmfgraph*}(110,60)
	   	\fmfleft{i,j}
        \fmfright{k,l}
      	\fmf{boson,label=$X$}{i,v1}
	   	\fmf{boson,label=$X$}{j,v1}
	   	\fmfv{decor.shape=circle,decor.filled=full,decor.size=5thick,
	   		l.d=10,
	   		label=$V^{h_iXX}_{\mu\nu}+\pthat V^{h_iXX}_{\mu\nu}$}{v1}
	   		\fmfv{decor.shape=circle,decor.filled=full,decor.size=5thick,
	   		l.d=10,
	   		label=$V^{h_iVV}_{\mu\nu}+\pthat V^{h_iVV}_{\mu\nu}$}{v2}
	   	\fmf{dashes,label=$\pthat\Delta_{ij}$}{v2,v1}
        \fmf{fermion,label=$V$,l.s=left}{k,v2}
	   	\fmf{fermion,label=$V$,l.s=left}{v2,l}
	\end{fmfgraph*}
\end{fmffile}
& \hspace{3.6cm}\begin{fmffile}{xxhhBI}
	\begin{fmfgraph*}(110,60)
	   	\fmfleft{i,j}
        \fmfright{k,l}
      	\fmf{boson,label=$X$}{i,v1}
	   	\fmf{boson,label=$X$}{j,v1}
	   	\fmfv{decor.shape=circle,decor.filled=full,decor.size=5thick,
	   		l.d=10,
	   		label=$V^{h_iXX}_{\mu\nu}+\pthat V^{h_iXX}_{\mu\nu}$}{v1}
	   	\fmfv{label=$V^{h}_{jkl}$,l.d=10}{v2}
	   	\fmf{dashes,label=$\pthat\Delta_{ij}$}{v2,v1}
        \fmf{dashes,label=$h_l$,l.s=left}{k,v2}
	   	\fmf{dashes,label=$h_k$,l.s=left}{v2,l}
	\end{fmfgraph*}
\end{fmffile}
\end{tabular}
\vspace{0.2cm}

\begin{tabular}{ccc}	        
\begin{fmffile}{xxhhQ}
	\hspace{0cm}\begin{fmfgraph*}(80,60)
	   	\fmfleft{i,j}
        \fmfright{k,l}
      	\fmf{boson,label=$X$,l.s=left}{i,v1}
	   	\fmf{boson,label=$X$,l.s=right}{j,v1}
        \fmf{dashes,label=$h_l$,l.s=right}{k,v1}
	   	\fmf{dashes,label=$h_k$,l.s=right}{v1,l}
	\end{fmfgraph*}
\end{fmffile}
& \hspace{0cm}
\begin{fmffile}{xxhhT}
	\hspace{1cm}\begin{fmfgraph*}(80,60)
	   	\fmfleft{i,j}
        \fmfright{k,l}
      	\fmf{boson,label=$X$,l.s=left}{i,v1}
	   	\fmf{boson,label=$X$,l.s=right}{j,v2}
	    \fmf{boson,label=$X$,l.s=right}{v2,v1}
        \fmf{dashes,label=$h_l$,l.s=right}{k,v1}
	   	\fmf{dashes,label=$h_k$,l.s=right}{v2,l}
	\end{fmfgraph*}
\end{fmffile} &
\begin{fmffile}{xxhhU}
	\hspace{1cm}\begin{fmfgraph*}(80,60)
	   	\fmfleft{i,j}
        \fmfright{k,l}
        \fmf{phantom}{v1,k}
        \fmf{phantom}{v2,l}
      	\fmf{boson,label=$X$,l.s=left}{i,v1}
	   	\fmf{boson,label=$X$,l.s=right}{j,v2}
	    \fmf{boson,label=$X$,tension=0.2,l.s=right}{v2,v1}
        \fmf{dashes,label=$h_k$,l.s=right,tension=0.2,l.d=20}{k,v2}
	   	\fmf{dashes,label=$h_l$,l.s=right,tension=0.2,l.d=20}{v1,l}
	\end{fmfgraph*}
\end{fmffile}
\end{tabular} 
\end{center}\vspace{-0.2cm}
\caption{\label{feyngraphsvvhh}The Feynman diagram for the
  Born-improved amplitudes $XX\rightarrow VV$ (with $V=Z,\,W$) and
$XX\rightarrow h_kh_l$.}
\end{figure}
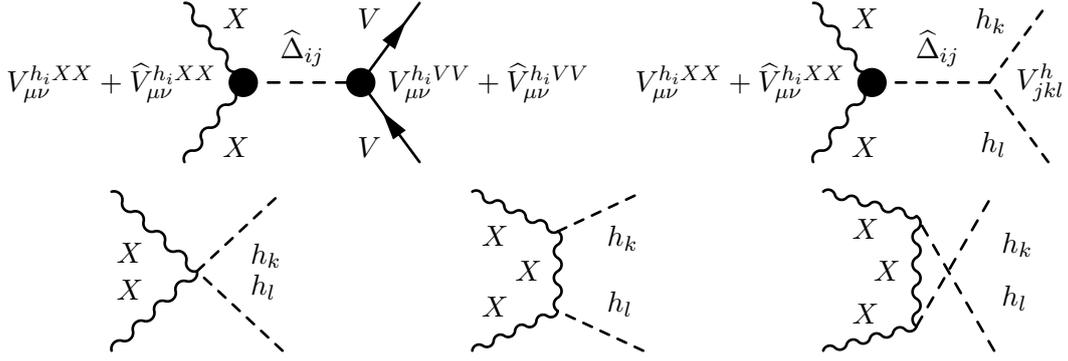

In addition to the resonant process $XX\rightarrow f\bar f$, there are
also the vector and scalar annihilation\- channels,
$XX\rightarrow (ZZ,\,W^+W^-)$ and $XX\rightarrow h_kh_l$, in the VDM
model, as~shown in fig. \ref{feyngraphsvvhh}. In order to obtain the
Born-improved amplitudes for these processes, we proceed as before. We
replace the tree-level propagator $\Delta^0(s)$ in the $s$-channel
graphs with the PT resummed propagator~$\pthat\Delta(s)$ given
in~(\ref{resprop}).  Likewise, we replace the tree-level couplings
with their
PT resummed counterparts~\cite{Papavassiliou:1997pb,Papavassiliou:1999qn}:
\begin{align}
  \label{mod_verhZZ}
V^{h_iZZ}_{\mu\nu}\ &\rightarrow\ \pthat \Gamma^{h_iZZ}_{\mu\nu}(q,p_1,p_2)\: =\:
V^{h_iZZ}_{\mu\nu}\: +\: \pthat V^{h_iZZ}_{\mu\nu}(q,p_1,p_2)\,,\\
  \label{mod_verhWW}
V^{h_iW^+W^-}_{\mu\nu}\ &\rightarrow\ \pthat \Gamma^{h_iW^+W^-}_{\mu\nu}(q,p_1,p_2)\: =\:
V^{h_iW^+W^-}_{\mu\nu}\: +\: \pthat V^{h_iW^+W^-}_{\mu\nu}(q,p_1,p_2)\,.
\end{align}
Note that the PT vertex corrections to $h_iZZ$ and $h_iW^+W^-$ satisfy tree-like Ward
identities analogous to those given in (\ref{WI1})--(\ref{WI4}). 

As can be seen from fig.~\ref{feyngraphsvvhh}, besides the resonant
$s$-channel graphs, there exist also non-resonant $t$- and $u$-channel
diagrams and four-point vertices contributing to the processes
$XX\rightarrow h_kh_l$ which do not require any improvement. The
high-energy behaviour of these amplitudes can be studied using the
GET. Because the structure of the vertex $V^h_{jkl}$ contains a piece
proportional to $R_{2j}$, there is no cancellation of the leading
terms in energy for both tree-level and Born-improved $s$-channel
diagrams as opposed to what happens for the process with $f\bar f$ in
the final state. Therefore, we expect that such graphs and the total
amplitude will go asymptotically to a constant ($\propto s^0$) at high
energies. One the other hand, if we replace the initial states $X$
with their respective Goldstone bosons~$G_X$, the presence of the
propagator suppresses the $s$-channel amplitude and makes the latter
fall as $s^{-1}$. Consequently, thanks to the ET, the high energy
behaviour of the total amplitude is dictated by the non-resonant part
of the amplitude with Goldstone bosons in the initial state,
i.e.~$G_XG_X \to h_k h_l$.  This part is not affected by resummation
and coincides with the tree-level result, as can easily be inferred
from the Feynman diagrams depicted in the second line of
fig.~\ref{feyngraphsvvhh}. This implies that for
$s/(4M_X^2)\rightarrow\infty$, the $s$-channel contribution to the
total Born-improved amplitude, which we denote here as
$\mathcal{\pthat A}^s_{X_LX_L\rightarrow h_k h_l}$, has to behave
exactly as its tree-level counterpart,
i.e.~$\mathcal{A}^s_{X_LX_L\rightarrow h_k h_l} \propto s^0$. Indeed,
this is the case, as the PT ensures that the classical
WI~(\ref{eq:WI3resum}) be satisfied at both tree and Born-improved
levels. Therefore, we obtain the same high-energy limit for the
Born-improved amplitude
$\mathcal{\pthat A}^s_{X_LX_L\rightarrow h_k h_l}$ as the one given by
the tree level amplitude, i.e.
\begin{equation}
   \label{eq:limampff}
\mathcal{\pthat A}^s_{X_LX_L\rightarrow h_k h_l}\
\overset{s\rightarrow\infty}=\ i\gx 
M_XR_{2i}\, \pthat\Delta^{-1}_{im}(s)\,
\pthat\Delta_{mj}(s)\,V^{h}_{jkl}\ =\ i\gx 
M_XR_{2i}V^{h}_{ikl}\ =\ {\rm const}\;.
\end{equation}
Observe that according to the ET, the high-energy limit of the total
Born-improved
amplitude${}$~$\mathcal{\pthat A}_{X_LX_L\rightarrow h_k h_l}$ will be
given by the contact diagram $G_XG_Xh_kh_l$ which tends to a constant
as $s/(4M_X^2)\to \infty$.



\section{Annihilation Cross-Sections and Evolution of Dark Matter Density}
\label{reldens}

In this section, we  apply the PT resummation approach  presented in the
previous section in order to compute the DM annihilation
cross-sections, as well as the evolution of the DM density within the VDM
model. To this end, we adopt the Born-improved amplitude${}$~\eqref{amp}
with the PT resummed propagator and one-loop dressed vertices for all the
annihilation\- channels, {\it viz}.
\begin{equation}
i\mathcal{\pthat A}^{XX\rightarrow \bar w w}_{\mu\nu} = \sum_{ij}
(V^{XXh_i}_{\mu\nu} + \pthat V^{XXh_i}_{\mu\nu})\, i\widehat\Delta_{ij\,}
\Gamma^{h_j \bar w w}\,,  
\end{equation}
where $w$ specifies the SM final state, i.e.~$w=Z,W,h_i,f$, and
$\Gamma^{h_j \bar w w}$ represents the proper $h_j\bar w w$-vertex
upon appropriate contraction with polarization vectors and
spinors~\footnote{Strictly speaking, one should include one-loop
  corrections to $h_i ZZ$- and $h_iW^+W^-$-vertices within the PT
  framework similarly to what was done in
  sec.~\ref{resummation}. However, these effects are found to be
  numerically negligible. Specifically, the one-loop corrections to
  $h_iXX$-vertex change the cross-sections by as much as~2~\%, when
  the dark gauge coupling $g_x$ approaches its perturbativity
  limit~$\sqrt{2\pi}$. The respective one-loop absorptive corrections
  to $h_i$ vertices with the SM $Z$ and $W^\pm$ vector bosons scale as
  $g^4$, so simple estimates indicate that these are at most at per
  mille level. Consequently, for most applications of phenomenological
  interest, the absorptive loop corrections to $h_i ZZ$- and
  $h_iW^+W^-$-vertices may safely be neglected.}. For the annihilation
channel into scalars (when $w=h_i$), we also include the non-resonant
contributions shown in fig.~\ref{feyngraphsvvhh}. The total
annihilation cross-section $\sigma$ is obtained as usual, by averaging
the squared amplitude over initial $X$-boson polarizations and summing
over all annihilation channels $w$ in the final state, their
polarizations and other possible internal degrees of
freedom. Furthermore, we have introduced in our analysis the parameter
\begin{equation} 
\delta\ \equiv\ \frac{4 \mx^2}{\mtwo^2}\: -\:1 \;,
\end{equation}
which provides a measure of proximity of the $h_2$-boson mass to the
DM $XX$ threshold.

\begin{figure}[!h]
\centering
\includegraphics[width=0.71\textwidth]{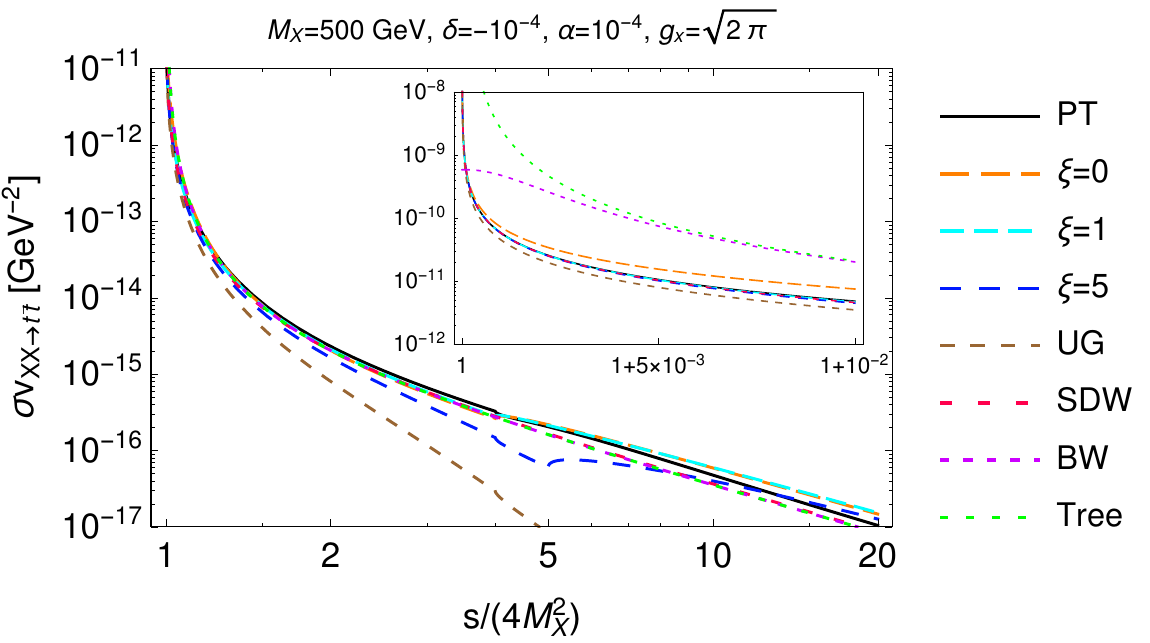}\\
\includegraphics[width=0.71\textwidth]{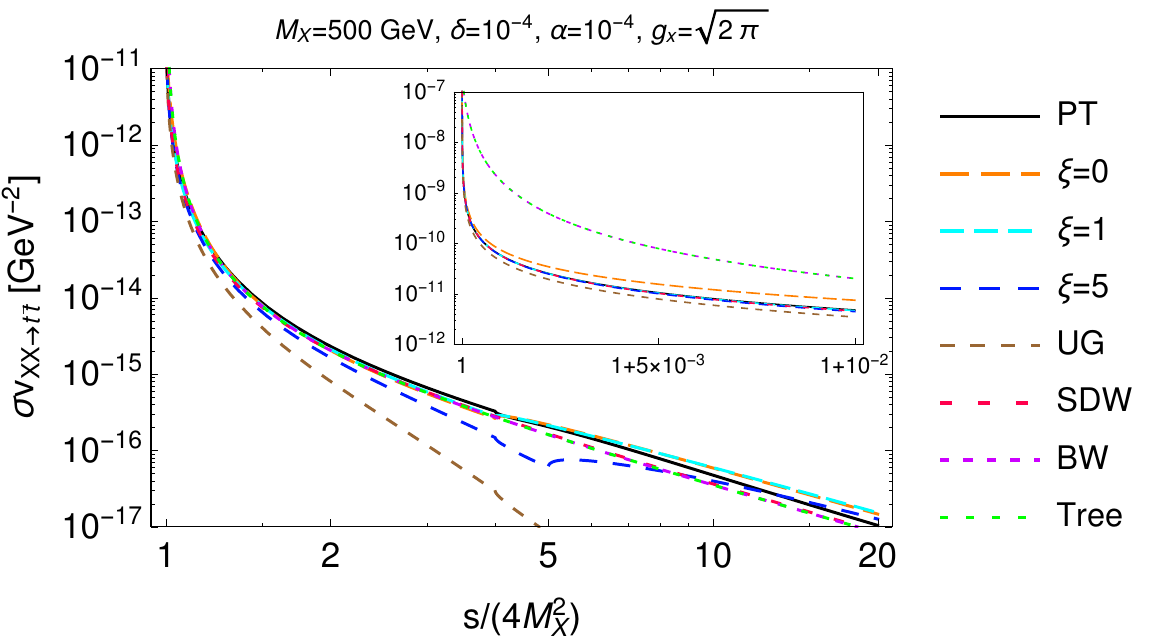}\\
\includegraphics[width=0.71\textwidth]{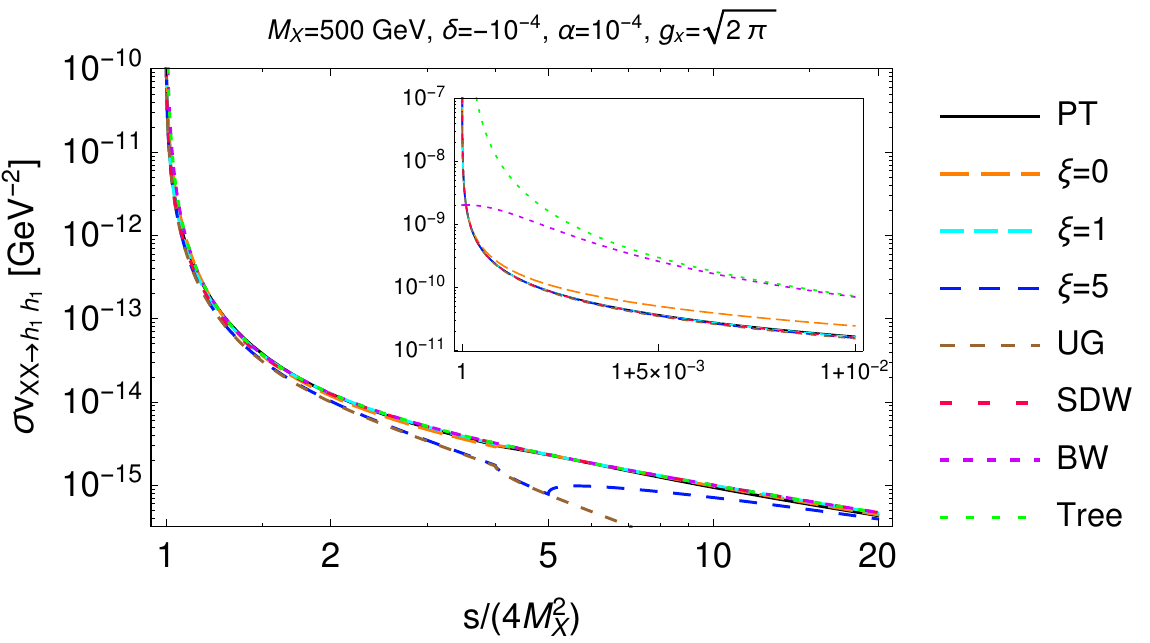}
\caption{\label{csplot}Resonant annihilation cross-section (times the
  relative velocity $v$) for $XX\rightarrow t\bar t$~(upper panels
  with negative and positive value of $\delta$)
  and $XX\rightarrow h_1 h_1$~(bottom panel) as a function of the
  energy ratio $s/(4M^2_X)$ obtained using the PT (solid line), a
  resummed propagator with self-energies calculated in the $R_\xi$
  gauge, for $\xi=0,\ 1,\ 5$, in the unitary gauge (UG), with the
  $s$-dependent Higgs width (SDW) as well as in the BW and tree-level
  approximation.  For the VDM model parameters, we set $M_X = 1$~TeV,
  $| \delta | = 10^{-4}$, $\alpha = 10^{-4}$ and $g_x = \sqrt{2\pi}$.
  Note that the inset plots display the respective cross-sections at
  the close vicinity of the DM threshold region.}
\end{figure}

Figure~\ref{csplot} shows the product $\sigma v$ of the annihilation
cross-section $\sigma$ with the relative velocity $v$ in the CoM frame
for the processes $XX\rightarrow t\bar t$ and $XX\rightarrow h_1h_1$,
as functions of the normalized energy squared parameter
\begin{equation}
   \label{eq:sbar}
\bar{s}\ \equiv\ \frac{s}{4M^2_X}\; .
\end{equation}
The results are obtained by utilising the PT resummed propagator, a
resummed BW propagator with self-energies calculated in the $R_\xi$
class of gauges with $\xi=0,\ 1,\ 5$ and in the unitary gauge~(UG). In
addition, fig.~\ref{csplot} displays the quantity $\sigma v$ computed
at the tree-level and in the BW approximation assuming a constant or
an $s$-dependent decay width~(SDW). In the BW case, we set
$\Gamma_i=\Im m\,\Pi_{ii}(m_i^2)/m_i$. In~the SDW approximation, we
include in the partial decay width $\Gamma_{h_2\rightarrow XX}$ the
exact form of the phase space factor corresponding to the $XX$
channel. This gives the leading energy-dependent correction to
$\Gamma_2$ near the $XX$ threshold, i.e.
\begin{equation}
   \label{eq:sdw}
\Gamma_{h_2\rightarrow XX}\ =\ 
\sqrt{\frac{1-4M_X^2/s}{1-4M_X^2/m_2^2}}\; \frac{\Im
  m\,\Pi^{(XX)}_{22}(m_2^2)}{m_2}\: \theta\big( s - 4M^2_X\big)\; . 
\end{equation}
For $\delta>0$, the above expression has to be analytically
continued. This SDW approximation works exceptionally well near the
Higgs pole, where~${s\approx m_2^2}$. Moreover, it does not modify the
high-energy behaviour of the tree-level amplitudes, because
$\Gamma_2(s)$ approaches a constant value when
$s/(4M_X^2)\rightarrow \infty$.

From fig.~\ref{csplot}, we can see that in the region where
$s/(4\mx^2)\simeq 1$, the PT result is close to the one that has been
evaluated in the $R_\xi$ gauges with $\xi=0,\,1,\,5$.  However, at large
$s$, the Born-improved (PT) cross-section (times the relative velocity
$v$) for the process $XX\to t\bar{t}$ varies between the one computed
in the $R_\xi$ gauge and that in the tree-level approximation, whereas
the cross-section calculated in the unitary gauge is suppressed. The
latter is a result of the $Z$- and $W^\pm$-boson contributions to
$\Pi_{ij}(s)$ which display a scaling behaviour
$\propto s^2/(16M_X^4)$ when the energy is well above the
corresponding $W^+W^-$ and $ZZ$ thresholds. For the annihilation
channel $XX\to h_1h_1$, the differences between the various results
are smaller, because the leading (in $s$) parts of the propagator do
not cancel asymptotically.  Moreover, the PT result approaches exactly
the tree-level one as discussed earlier. Observe that near the
threshold the naive BW and tree-level results deviate dramatically
from those found in the~PT, in contrast to the approximation with
$s$-dependent $XX$ phase space factor~\eqref{eq:sdw} which leads to a
very good agreement in this region. For $\delta<0$, the large width
$\Gamma_2$ used in the BW approximation results in an underestimated
cross-section right at the threshold. On the other hand, for
$\delta>0$, the width $\Gamma_2$ contains only SM contributions which
are suppressed by the small mixing angle $\alpha$. For this reason, the BW
approximation is close to the tree-level result. Finally, we note the
presence of the fictitious threshold for $\xi=5$. This is an example
of a gauge artifact that originates from a naive resummation of
self-energies in the $R_\xi$~class of gauges.

\begin{figure}[!ht]
\centering
\includegraphics[width=0.75\textwidth]{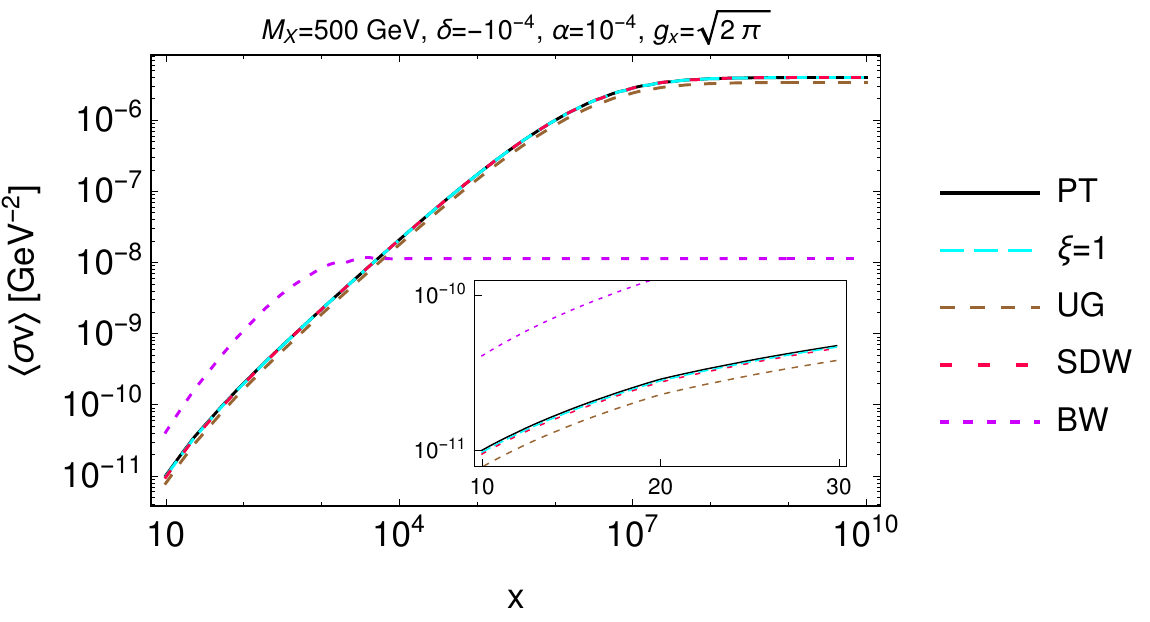}\\
\includegraphics[width=0.75\textwidth]{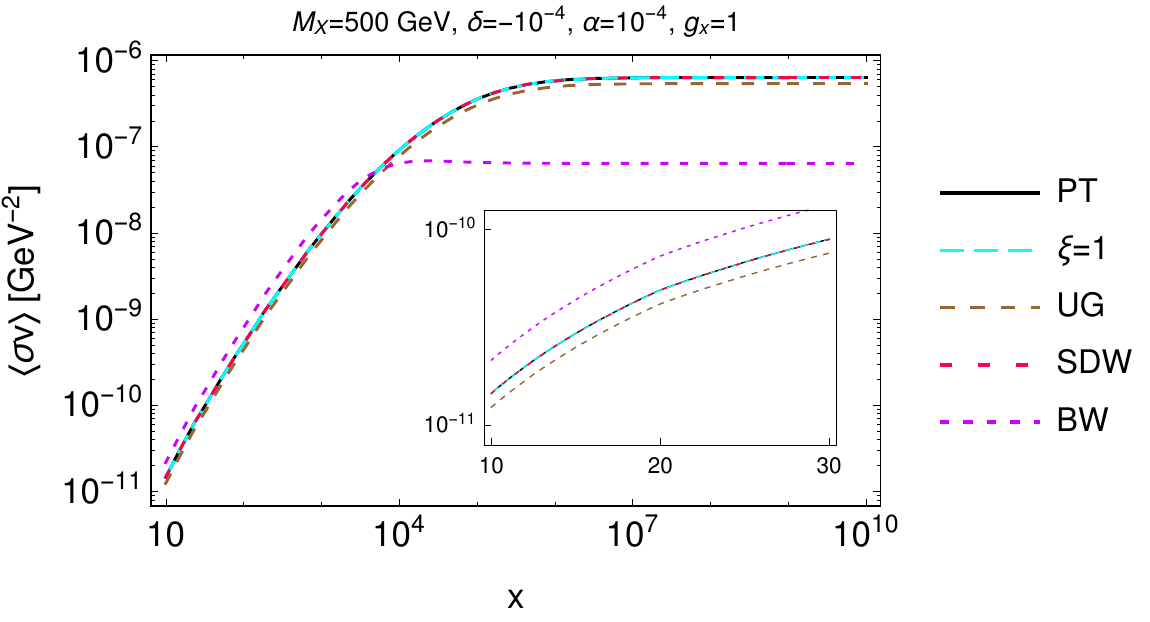}\\
\includegraphics[width=0.75\textwidth]{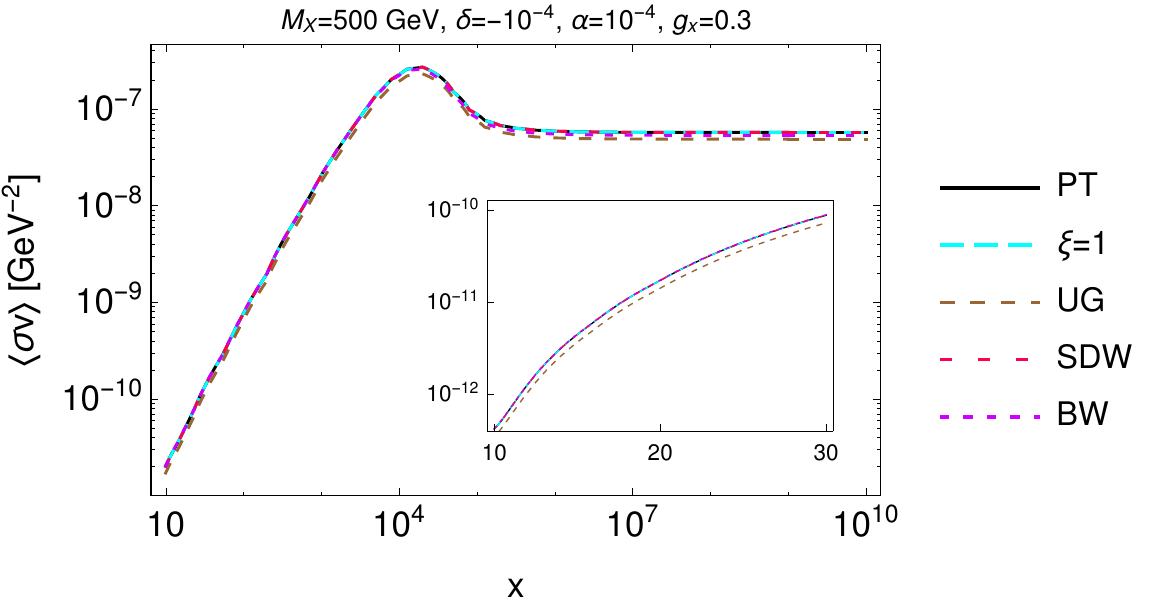}
\caption{\label{sigvplot}Thermally averaged total annihilation
  cross-section $\sigv(x)$ as a function of $x\equiv M_X/T$, for
  $g_x=0.3$, $1$ and $\sqrt{2\pi}$ and the remaining model parameters
  same as in fig.~\ref{csplot}. The solid line corresponds to the
  cross-section (\ref{sigmav}) obtained using the PT with resummed
  propagator and corrected $h_iXX$ vertex while the dashed lines are
  for the self-energies calculated using $R_\xi$ gauges for several
  values of the gauge parameter $\xi$, the unitary gauge (UG),
  standard BW approximation with constant (BW) or s-dependent widths
  (SDW) given by (\ref{eq:sdw}).}
\end{figure}

The evolution of DM density is mainly governed by the size of the
thermally averaged annihilation cross-section $\sigv$, which may be 
computed as a function of $x\equiv M_X/T$ by employing the standard
formula~\cite{Gondolo:1990dk}:
\begin{equation}
   \label{sigmav}
\sigv\ =\ \frac{2x}{K_2^2(x)} \int_{1}^{\infty}d\bar{s} \; \sigma v \;
\sqrt{\bar{s} - 1}\, \bar{s}\, K_1(2x\sqrt{\bar{s}})\ , 
\end{equation}
where the dimensionless parameter $\bar{s}$ is defined
in~\eqref{eq:sbar} and $K_n(x)$ denotes the modified Bessel function
of order $n$.  In fig.~\ref{sigvplot}, we present the values of
$\sigv$ for $g_x=0.3$, $1$ and $\sqrt{2\pi}$~\footnote{The value
  $\gx=\sqrt{2\pi}$ corresponds to the saturation of the
  perturbativity bound for the quartic coupling
  $\lambda_{H,S} \le 4\pi$ for the resonant region of the VDM model,
  for which $2M_X\approx m_2$.}.  The results were calculated in the
PT, the Feynman gauge ($\xi=1$), the unitary gauge (UG) and in the BW
approximation with the usual constant decay widths and $s$-dependent
widths (SDW) as stated in~\eqref{eq:sdw}.  We see that as the dark
gauge coupling $g_x$ decreases, the standard BW approximation starts
to describe more accurately the resonant cross-section. This follows
from the fact that $XX$ contribution to the self-energy $\Pi_{ij}(s)$,
which strongly depends on the energy close to the threshold, ceases to
dominate over the nearly constant contributions from the SM
states. Then, as presented in the lower panel of fig.~\ref{sigvplot}
for $g_x=0.3$, the thermally averaged cross-section $\sigv$ forms a
hilltop located at $x=1/|\delta|$. This is the value of the
temperature for which the thermal distribution (\ref{sigmav}) is centered at the
resonance peak $s=m_2^2$.

\begin{figure}[!htb]
\centering
\includegraphics[width=0.43\textwidth]{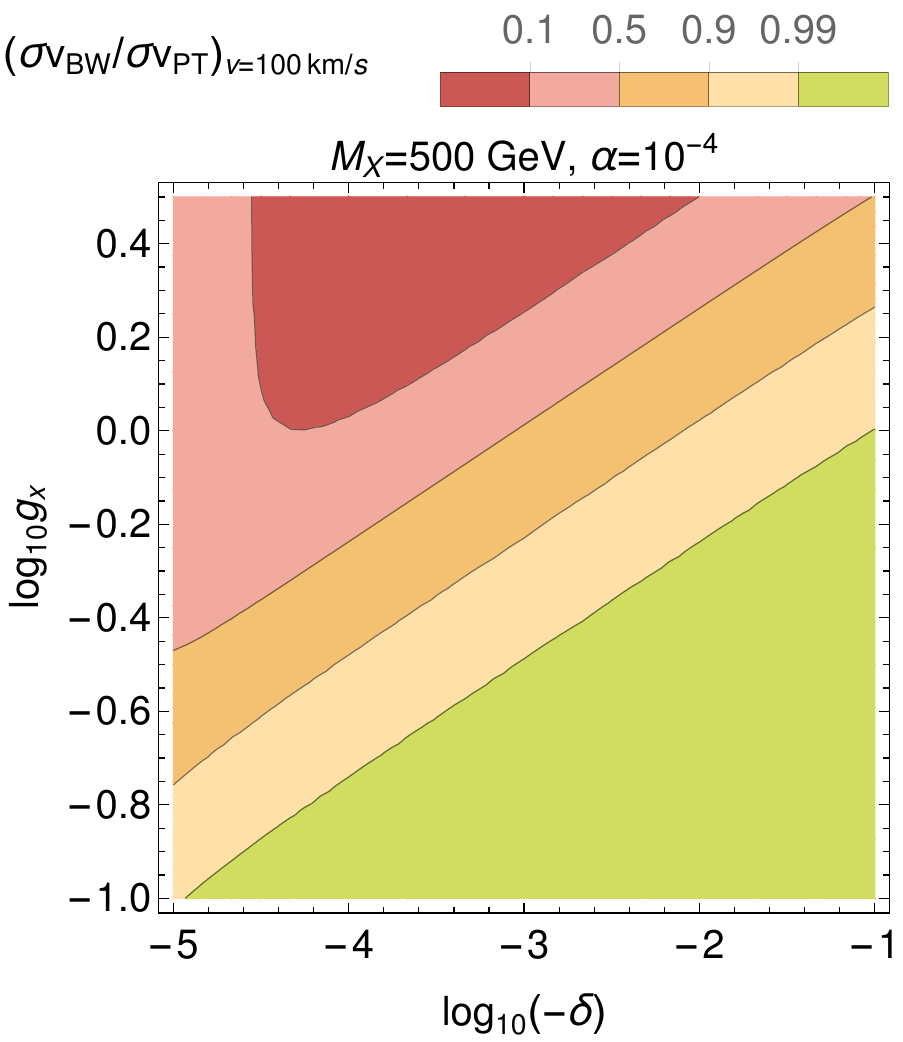}\hspace{0.5cm}
\includegraphics[width=0.43\textwidth]{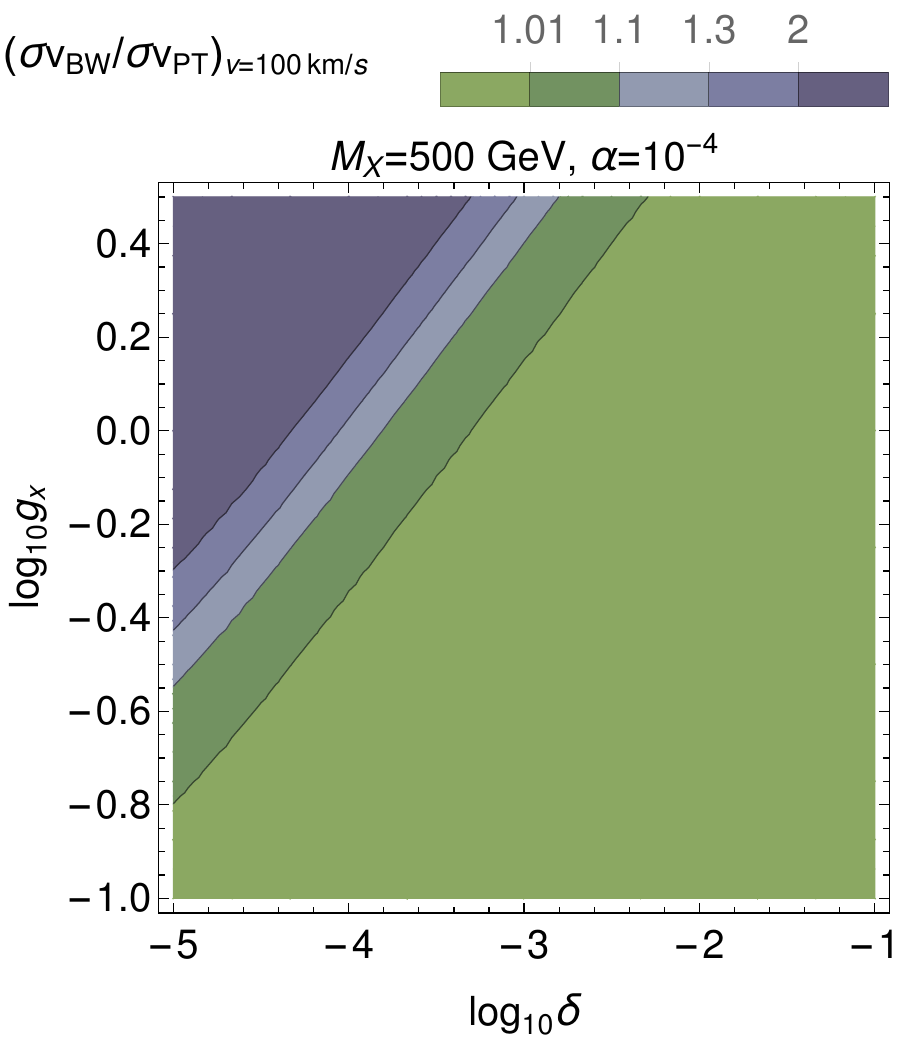}\\
\hspace{-0.2cm}\includegraphics[width=0.46\textwidth]{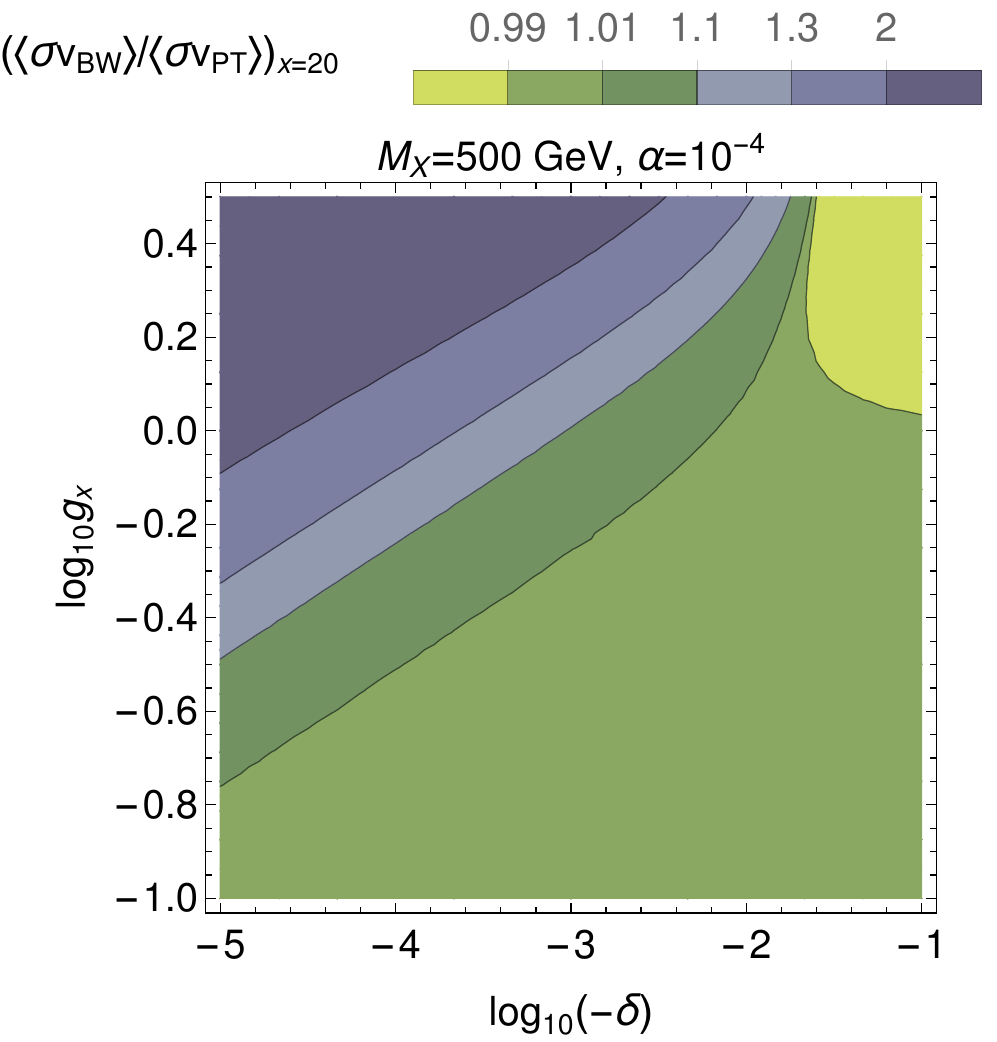}\hspace{0.3cm}
\includegraphics[width=0.43\textwidth]{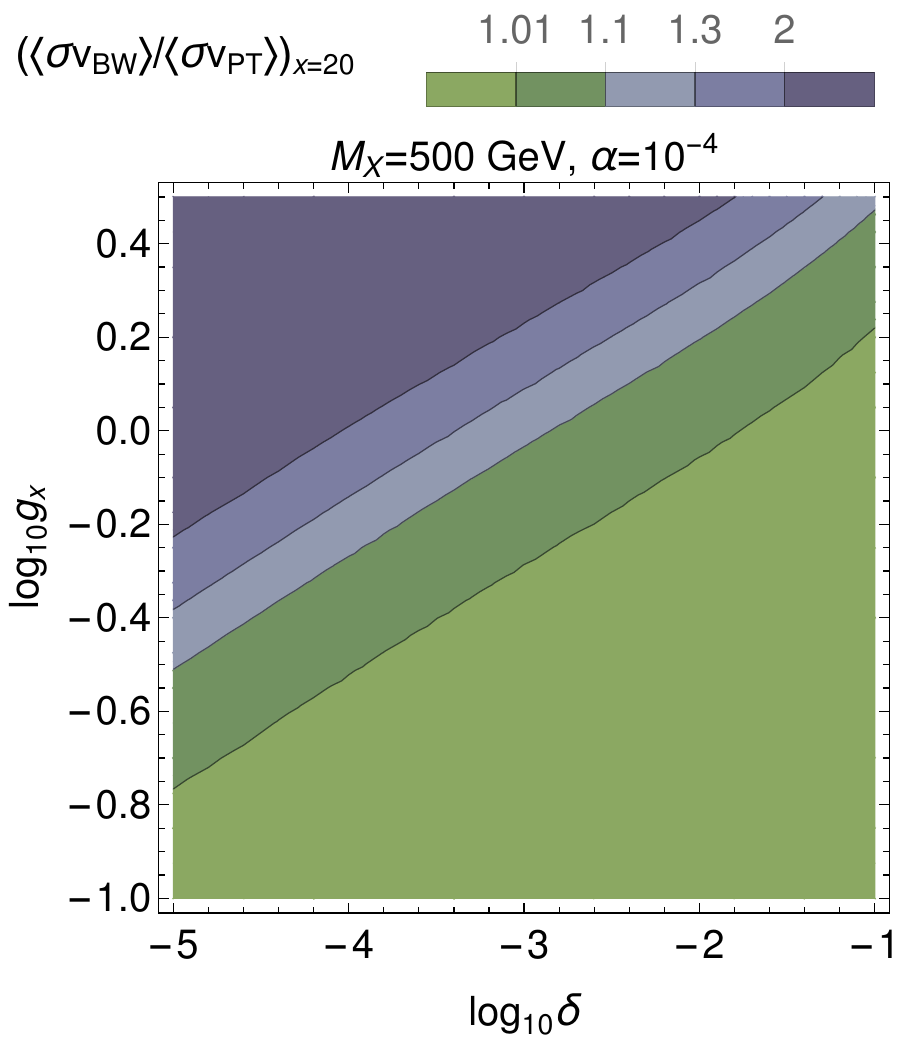}\\
\caption{\label{compcs}Numerical comparisons in the plane ($\delta$, $\gx$) of
  the cross-sections obtained using standard BW approximation and PT
  method. In the upper panels contours of $\sigma v_{BW}/\sigma v_{PT}$
  for $v=100$~km/s are plotted, while in the bottom panels contours of
  $\langle \sigma v_{BW} \rangle / \langle \sigma v_{PT} \rangle$ at
  $x=20$ are shown. The panels on the LHS and RHS present results for $\delta<0$ and $\delta>0$ respectively.}
\end{figure}

In fig. \ref{compcs}, we display numerical comparisons of thermally
averaged cross-sections computed by means of the PT and the standard
BW approximation on the plane defined by the parameters $g_x$ and
$\delta$. We observe that the BW approximation is applicable only if
the parameters $g_x$ and $|\delta|$ are sufficiently small. Otherwise,
its simplistic use for relative velocities $v=100$~km/s (typical in astrophysical searches of DM annihilation signal) can lead to
annihilation cross-sections that are much smaller (larger) than the PT
result, for negative (positive) values of $\delta$. On the other hand,
for thermally averaged cross-sections calculated at temperatures close
to that of chemical decoupling ($x=20$), the BW approximation gives
results that are comparable to or larger than those
derived by the PT resummation method.

In the following, we will analyze the evolution of the relic DM
density. Following the recent
studies~\cite{Duch:2017nbe,Binder:2017rgn}, we take into consideration
the effect of early kinetic decoupling, which turns out to be an
important phenomenon for scenarios with resonant DM annihilation. This
effect appears, because the resonantly enhanced annihilation rate
corresponds to scattering processes of suppressed strength, whose role
is to maintain the kinetic equilibrium\- between the DM and the
thermal${}$~bath in the early Universe. Moreover, as readily observed
from fig.~\ref{sigvplot}, the thermally averaged cross-section $\sigv$
grows by many orders of magnitude when $x$ varies from the chemical
decoupling temperature of $x\approx 20$ until $x \sim {\cal O}(10^4)$.
Therefore, the effective DM annihilation can be prolonged to a period
when the DM does no longer have the same temperature as that of the SM
thermal~bath.

In order to determine the evolution of DM density, we assume that the
DM distribution does not deviate significantly from the thermal one.
Therefore, we may describe it with the DM temperature $T_X$. To
compute the DM relic abundance, we solve numerically a set of coupled
Boltzmann equations for the DM yield $Y$ and the temperature
parameter~$y$~\cite{vandenAarssen:2012ag,Binder:2017rgn}
\begin{align}
   \label{eq:boltzdens}
\frac{dY}{dx}\ =&\ -\frac{s}{x H}\bigg(1+\frac{T}{3h}\frac{dh}{dT}\bigg) 
\bigg(Y^2\sigv_{T=T_X} - Y_{\rm eq}^2\sigv \
                  \bigg)\ ,\\[7pt]
   \label{eq:boltztemp}
\frac{dy}{dx}\ =&\ -\frac{\gamma(T)}{xH}\Big(1+\frac{T}{3h}\frac{dh}{dT}\Big)
  \bigg(y-y_{\rm
  eq}\bigg)+\bigg(1+\frac{T}{3h}\frac{dh}{dT}\bigg)\frac{\langle
  p^4/E^3\rangle_{T=T_X}}{3xT_X} \\[5pt] 
&\ -\frac{sy}{xHY}\bigg(1+\frac{T}{3h}\frac{dh}{dT}\bigg)
  \bigg[Y^2\bigg(\sigv_{2}-\sigv\bigg)_{T=T_X}-Y_{\rm
  eq}^2\bigg(\frac{y_{\rm eq}}{y}\sigv_2-\sigv\bigg)\bigg]\, ,\non
\end{align}
where
\begin{align}
&Y=\frac{n_X}{s}\ ,  \;\;\;\; y = \frac{M_X T_X}{s^{2/3}}\ ,\;\;\;\;\;
T_{X,{\rm eq}} = T\ ,\;\;\;\;\; n_{X,{\rm eq}}=\frac{g_X M_X^3}{2\pi^2 x} K_2(x)\ ,\non \\[5pt]
&H=\sqrt{\frac{4\pi^3 g}{45 m^2_{\rm Pl}}}T^2 \ ,\;\;\;\;\; s=\frac{2\pi^2}{45}h T^3\ .
\end{align}
In the above, $m_{\rm Pl}$ is the Planck mass, $n_X$ denotes the number
density of DM, $s$ is the entropy density, $H$ is the Hubble
parameter, $g_X=3$, whereas $g$ and $h$ are respectively the effective
numbers of relativistic degrees of freedom for energy density and
entropy. Besides the collision terms $\propto \sigv$, the coupled
equations (\ref{eq:boltzdens}) and (\ref{eq:boltztemp}) contain also
the following thermal averages:
\begin{equation}
\sigv_2 =\int_{1}^{\infty}d\bar{s} \sigma v
\frac{4x^3\bar{s}^{2}}{3K_2^2(\bar{s})}\int_{1}^\infty d\epsilon_+
e^{-2\epsilon_+ \sqrt{\bar{s}}}\bigg[\epsilon_+z
+\frac{1}{2\sqrt{\bar{s}}}\log\bigg(\frac{\sqrt{\bar{s}}\epsilon_+ -
    z}{\sqrt{\bar{s}}\epsilon_+ + z}\bigg)\bigg]\ , 
\label{sigmav2}
\end{equation}
\begin{equation}
\langle p^4/E^3\rangle=\frac{g_X}{2\pi^2 n_{X,{\rm eq}}(T)}\int dp \frac{p^6}{E^3}e^{-E/T}\, ,
\end{equation}
where $E$ and $p$ are the energy and momentum of the DM and $z\equiv
\sqrt{(\bar{s}-1)(\epsilon_+^2-1)}$. Finally, the scattering momentum
exchange rate $\gamma(T)$ may be expressed as 
\begin{equation}
\gamma(T)=\frac{1}{3\pi^2g_XM_X}\sum_{\rm SM}\int_{m_{\rm SM}}^\infty d\omega\,
g^\pm (\omega)\,\partial_\omega\big[(\omega^2-m^2_{\rm SM})^2\sigma_T|_{s=M_X^2+2\omega M_X+m_{\rm SM}^2}\big] \; , 
\end{equation}
where $m_{\rm SM}$ is the mass of the SM state upon which the DM scatters,
$g^\pm(\omega)=1/[\exp(\omega/T)\pm 1]$ denotes its phase-space
distribution and $\sigma_T=\int d\Omega\,
(1-\cos\theta\,d\sigma/d\Omega)$ is the standard transfer
cross-section for elastic scattering. The sum runs over all
relativistic degrees of freedom present in the SM thermal bath.	

\begin{figure}[!ht]
 \centering
 \includegraphics[width=0.89\textwidth]{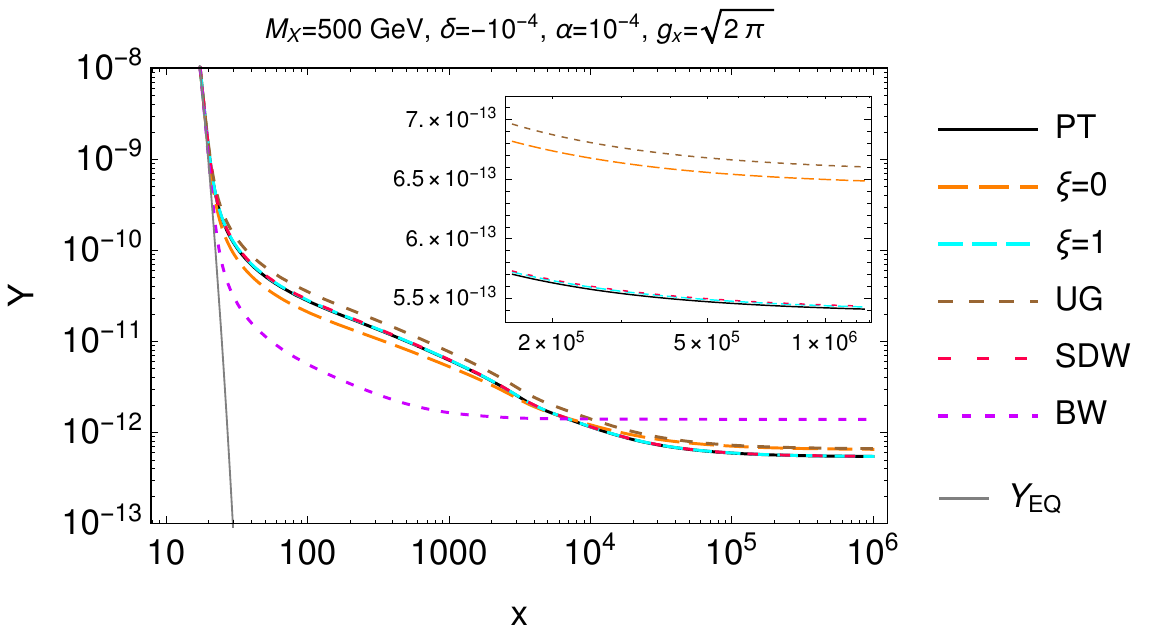}\\
  \includegraphics[width=0.89\textwidth]{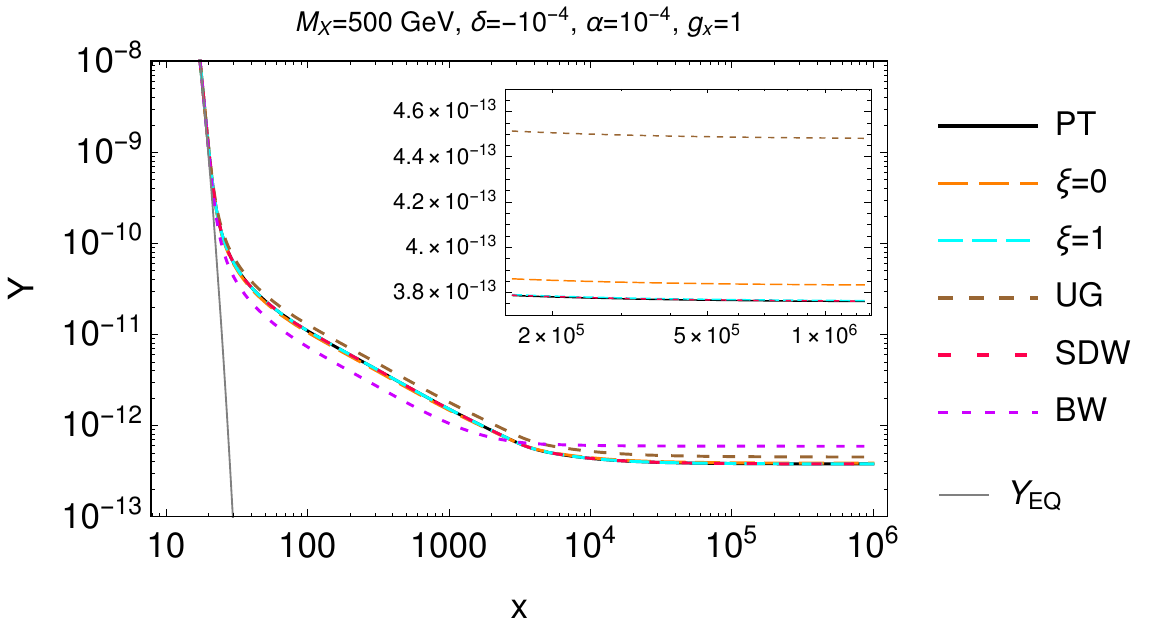}\\
 \caption{Dark matter yield $Y(x)$ for $\gx=\sqrt{2\pi}$ (upper panel)
   and $\gx=1$ (lower panel) obtained using the PT, $R_\xi$ gauge
   ($\xi=0,\,1$), unitary gauge (UG), the standard Breit-Wigner
   propagator or approximate $s$-dependent width (SDW).} 
 \label{Yfull}
\end{figure}

The different predictions for the thermally averaged
cross-sections $\sigv$ calculated using various methods manifest
themselves in different evolutions of the DM yield $Y(x)$. As can be
seen from fig.~\ref{Yfull},  the result obtained in the
unitary gauge or in the Landau gauge~${\xi=0}$ deviates
significantly from the PT evaluation. Likewise, the standard BW
approximation completely fails to describe the underlying dynamics if
$\gx=\sqrt{2\pi}$. Instead, an evaluation in the usual
Feynman gauge~${\xi=1}$ (where no unphysical thresholds occur) or in an
$s$-dependent width approximation for the BW propagator of the resonant
mediator turns out to be both very good approximations of the PT
resummation approach.

\section{Conclusions}
\label{con}

In processes of resonant DM annihilation, one central difficulty
arises from the use of a BW approximation with a constant width for
the propagator of the exchanged particle in the $s$-channel. In
particular, if the respective DM channel contributes significantly to
the decay width of this mediator particle, the evaluation of the
resonant transition amplitude can become very inaccurate close to the
DM production threshold.  This problem may be partially circumvented
by utilising a running $s$-dependent width for the
mediator~\cite{Duch:2017nbe}, which results from the imaginary
(absorptive) part of the mediator's self-energy. However, in
spontaneously broken gauge theories, such a running width becomes
gauge-dependent in the off-shell region.

In this paper, we addressed this issue within the gauge-independent
framework of the Pinch Technique (PT). The PT has many desirable
field-theoretic properties, including analyticity, unitarity and the
gauge invariance of the classical action.  As an illustrative
scenario, we considered the SM extended by a local U(1)$_X$ symmetry
which may lead to a massive stable gauge field $X$ that could
successfully play the role of a Vector DM. By adopting a
resummation approach implemented by the PT, we calculated the PT
resummed propagator for the scalar $h_ih_j$ system (with $i,j =1,2$).
In addition, we calculated the relevant one-loop corrections to
the $h_iXX$-coupling in the PT. As we have shown in this work, the
latter are necessary in order to obtain a transition amplitude for
DM-pair annihilation which has the proper high-energy limit in
agreement with both the Equivalence Theorem and the Generalized
Equivalence Theorem.

The PT resummation method enabled us to derive a Born-improved
amplitude which was used to obtain self-consistent and accurate
predictions for the DM abundance of the $X$ vector boson. We have
illustrated how these predictions differ from those that one finds
using other methods of resummation. In our comparative analysis, we have
taken into account the effect of early kinetic decoupling of the
dark-sector particles from the SM thermal bath. We have shown that the
standard BW approximation is not applicable to large part of the
parameter space in which the DM contribution to the mediator's
self-energy dominates over the SM one. Its mere use results not only
in distorted predictions for the DM relic abundance, but also in wrong
annihilation cross-sections for indirect detection experiments. To
deal with this problem, we devised an alternative simple approximation
that utilises an energy-dependent width which yields the correct DM
annihilation rates and relic abundance.  In~this~study, we presented
results for resonant DM annihilation processes only. Nevertheless, it
is straightforward to extend our considerations to DM elastic
scatterings. We postpone such an analysis to a future communication.

\section*{Acknowledgments}
BG acknowledges support by the National Science Centre (Poland),
research projects no~2014/15/B/ST2/00108, no~2017/25/B/ST2/00191. MD
acknowledges support by the National Science Centre (Poland), research
project no~2017/25/N/ST2/01312. Finally, the work of AP is supported in part by the
Lancaster--Manchester--Sheffield Consortium${}$ for Fundamental
Physics, under STFC research grant ST/L000520/1.


\appendix

\section{The Abelian Vector Dark Matter Model}
\label{model}

The simplest model of vector DM (VDM) is an extension of the SM by an
extra $\uone$ gauge group together with a complex scalar $S$ that is
charged under~$\uone$~\cite{Hambye:2008bq,Lebedev:2011iq,Farzan:2012hh,Baek:2012se,
  Baek:2014jga,Duch:2015jta}. The extra scalar $S$ is singlet under
the SM gauge group and has non-zero VEV, and as such it provides, via
the standard Higgs mechanism, mass for the $\uone$ vector boson $X_\mu$
which plays here the role of the VDM. Its stability can be ensured by a
`dark' charge symmetry which acts in the hidden DM sector as
follows: 
\begin{equation} 
X_{\mu}\ \rightarrow\ -\;X_{\mu} \;, \qquad 
S\ \rightarrow\ S^*\;. 
\end{equation}
In the non-linear representation of the complex scalar field 
$S=\phi e^{i\sigma}$, the `dark' charge symmetry implies that 
$\phi\rightarrow \phi$ and $\sigma\rightarrow -\sigma$,
whereas all other fields are neutral under this symmetry.

In this VDM model, the VEVs of the scalar fields are given by
$(\langle H \rangle,\langle S \rangle)=\frac{1}{\sqrt{2}}(v,\vx)$
which generate the gauge-boson masses
\begin{equation} 
M_W = \inv2 g v\; , \qquad 
M_Z = \frac{1}{2}\sqrt{g^2+g'^2} v\;, \qquad 
\mzp = \gx \vx\;,
\end{equation} 
where $g$, $g'$ and $g_x$ are the SU(2)$_L$, U(1)$_Y$ and
U(1)$_X$ gauge coupling constants, respectively. 

The scalar potential of the VDM model reads 
\begin{equation} 
V\ =\ -\mu^2_H|H|^2 +\lambda_H |H|^4 -\mu^2_S|S|^2
+\lambda_S |S|^4 +\kappa |S|^2|H|^2\; ,
\end{equation} 
with $H=\big( H^+\,,\, H^0\big)^{\sf T}$.
To determine the physical scalar spectrum, we expand the scalar fields
linearly about their respective VEVs as follows:
\begin{equation}
S=\frac{1}{\sqrt{2}}(\vx+ \phi_S +iG_X) \ \ , \ \ H^0=
\frac{1}{\sqrt{2}}(v + \phi_H+ iG_Z)\;.  
\end{equation} 
Our next step is to diagonalise the square mass matrix $\mathcal{M}^2$
for the fluctuations $ \left(\phi_H, \phi_S\right)$. This can be done
by carrying out an orthogonal O(2) rotation $R$ which is usually
parameterised by a mixing angle $\alpha$, such that
$\mathcal{M}_{\text{diag}}^2= R^{-1} \mathcal{M}^2 R$. In this way, we
obtain
\begin{equation}
\mathcal{M}_{\text{diag}}^2 = \left(
\begin{array}{cc}
\mone^2  &0 \\ 
0 &\mtwo^2
\end{array} 
\right)\; ,
 \qquad
R = \left( 
\begin{array}{cc}
\cos \alpha   & -\sin \alpha \\ 
\sin \alpha &\cos \alpha
\end{array} 
\right)\;, 
 \qquad
 \left( 
\begin{array}{c}
h_1\\ 
h_2
\end{array} 
\right) = R^{-1}
\left(
\begin{array}{c}
\phi_H\\ 
\phi_S
\end{array} 
\right)\,,
\end{equation} 
where $h_1$ is the observed Higgs particle with a mass
$m_1=125\gev$. All scalar vertices of the theory may be described in
terms of the four extra parameters: $M_X$, $m_2$, $g_x$ and $\alpha$,
in addition to the SM ones. Specifically, the quartic couplings of the
VDM potential may be expressed as follows:
\begin{equation} 
\lambda_h=g^2\frac{m_1^2 c^2_\alpha+m_2^2
  s^2_\alpha}{8M_W^2}\ ,\;\;\;\;\; \lambda_s=\gx^2\frac{m_1^2
  s^2_\alpha+m_2^2c^2_\alpha}{8M_W^2}\ ,\;\;\;\;\;
\kappa=g\gx\frac{(m_1^2-m_2^2)s_{2\alpha}}{4M_W M_X}\ , 
\end{equation} 
with $c_\alpha\equiv \cos\alpha$ and $s_{\alpha}\equiv \sin\alpha$.

\vfill\eject
\section{Feynman Rules in the $R_\xi$ and Covariant Background
  Field Gauges} 
\label{bfgfeyn}
\begin{table}[h]
\begin{center}
\tetab
\setlength{\tabcolsep}{9pt} 
\begin{tabular}{|c|c|c|c|}
\hline
& &    & \vspace{-0.5cm} 
\\  
\begin{fmffile}{ZpZphi}
	        \begin{fmfgraph*}(60,50)
	            \fmfleft{i,j}
	            \fmfright{k}
	            \fmf{boson,label=$\zp$}{i,v}
		    \fmf{boson,label=$\zp$}{j,v}
		    \fmf{dashes,label=$h_i$}{v,k}
	        \end{fmfgraph*}
\end{fmffile} 
&
 \begin{fmffile}{ZZhi}
	        \begin{fmfgraph*}(60,50)
	            \fmfleft{i,j}
	            \fmfright{k}
	            \fmf{boson,label=$Z$}{i,v}
		    \fmf{boson,label=$Z$}{j,v}
		    \fmf{dashes,label=$h_i$}{v,k}
	        \end{fmfgraph*}
\end{fmffile}  & 
 \begin{fmffile}{WWhi}
	        \begin{fmfgraph*}(60,50)
	            \fmfleft{i,j}
			\fmfright{k}
			\fmf{boson,label=$W^+$}{i,v}
			\fmf{boson,label=$W^-$}{j,v}
			\fmf{dashes,label=$h_i$}{v,k}
		  \end{fmfgraph*}
  \end{fmffile} &
  \begin{fmffile}{FFhi}
		  \begin{fmfgraph*}(60,50)
		      \fmfleft{i,j}
		      \fmfright{k}
		      \fmf{fermion,label=$\bar{f}$}{i,v}
		      \fmf{fermion,label=$f$}{j,v}
		      \fmf{dashes,label=$h_i$}{v,k}
		  \end{fmfgraph*}
  \end{fmffile}
\\
& &    & \vspace{-0.7cm} 
\\
$i2\gx\mzp R_{2i}$
& $ig \frac{M_Z^2}{M_W} R_{1i}$
& $ig M_W R_{1i}$
& $ig {m_f}{2 M_W} R_{1i}$
\\\hline
\multicolumn{3}{|c|}{\hspace*{-2.5cm} \vbox{\vspace{-0.2cm}\begin{equation*}
\begin{split}
 V^h_{ijk}&= i[\kappa v (R_{1i} R_{2j} R_{2k} + R_{2i} R_{1j} R_{2k} + R_{2i} R_{2j} R_{1k}) \\ 
 &+ \kappa \vx (R_{2i} R_{1j} R_{1k} + R_{1i} R_{2j} R_{1k} + R_{1i} R_{1j} R_{2k})\\
 &+ 6\lambda v ( R_{1i} R_{1j} R_{1k} ) + 6 \lambda_s \vx (R_{2i} R_{2j} R_{2k})]
 \end{split}
\end{equation*}\vspace{-0.4cm}}\hspace*{-2.5cm}} & 
 \begin{fmffile}{hihjhk}
	        \begin{fmfgraph*}(60,50)
	            \fmfleft{i,j}
	            \fmfright{k}
	            \fmf{dashes,label=$h_i$}{i,v1}
		    \fmf{dashes,label=$h_j$}{j,v1}
		    \fmf{dashes,label=$h_k$}{k,v1}
	        \end{fmfgraph*}
\end{fmffile}
\\\hline
\vspace{-0.5cm}
& & & 
\\
\begin{fmffile}{goldh}
	        \begin{fmfgraph*}(60,50)
	            \fmfleft{i,j}
	            \fmfright{k}
	            \fmf{dashes,label=$G_\zp$,l.s=right}{i,v1}
		    \fmf{dashes,label=$G_\zp$,l.s=left}{j,v1}
		    \fmf{dashes,label=$h_i$}{k,v1}
	        \end{fmfgraph*}
\end{fmffile}
& 

 \begin{fmffile}{cchi}
		  \begin{fmfgraph*}(60,50)
		      \fmfleft{i,j}
		      \fmfright{k}
		      \fmf{dashes_arrow,label=$\bar{c}$,l.s=left}{v,i}
		      \fmf{dashes_arrow,label=$c$}{j,v}
		      \fmf{dashes,label=$h_i$}{v,k}
		  \end{fmfgraph*}
  \end{fmffile}

&
 \begin{fmffile}{hhZpZp}
		  \begin{fmfgraph*}(60,50)
		      \fmfleft{i,j}
	            \fmfright{k,l}
	            \fmf{boson,label=$\zp$,l.d=20}{v1,i}
		    \fmf{boson,label=$\zp$,l.d=20}{j,v1}
		    \fmf{dashes,label=$h_i$,l.d=20}{k,v1}
		    \fmf{dashes,label=$h_j$,l.d=20}{v1,l}
	        \end{fmfgraph*}
\end{fmffile}
&
 \begin{fmffile}{goldhw}
	        \begin{fmfgraph*}(60,50)
	            \fmfleft{i,j}
	            \fmfright{k}
	            \fmf{dashes,label=$G_\zp$,l.s=left,label.dist=10}{v1,i}
		    \fmf{dashes,label=$h_i$,label.dist=10}{j,v1}
		    \fmf{boson,label=$\zp$}{k,v1}
		    \marrow{a}{down}{bot}{$k$}{i,v1}
		    \marrow{b}{up}{top}{$p$}{j,v1}
	        \end{fmfgraph*}
\end{fmffile}
\\
$-i \gx \frac{m^2_i}{\mzp}R_{2i}$ & 
$-i \gx \mzp \xi_X$ &
$i2\gx^2 R_{2i}R_{2j}$ &
$-g(k_\mu-p_\mu)R_{2i}$ 
\\\hline
\vspace{-0.5cm}
& & & 
\\
\begin{fmffile}{goldxx}
	        \begin{fmfgraph*}(60,50)
	            \fmfleft{i,j}
	            \fmfright{k,l}
	              \fmf{dashes,label=$G_\zp$,l.s=left,l.d=2}{i,v1}
		    \fmf{dashes,label=$G_\zp$,l.s=right,l.d=2}{j,v1}
		    \fmf{boson,label=$\zp$,l.s=right,l.d=2}{k,v1}
		    \fmf{boson,label=$\zp$,l.d=2}{v1,l}
	        \end{fmfgraph*}
\end{fmffile}

& \begin{fmffile}{goldgg}
	        \begin{fmfgraph*}(60,50)
	            \fmfleft{i,j}
	            \fmfright{k,l}
	              \fmf{dashes,label=$G_\zp$,l.s=left,l.d=2}{i,v1}
		    \fmf{dashes,label=$G_\zp$,l.s=right,l.d=2}{j,v1}
		    \fmf{dashes,label=$G_\zp$,l.s=right,l.d=2}{k,v1}
		    \fmf{dashes,label=$G_\zp$,l.d=2}{v1,l}
	        \end{fmfgraph*}
\end{fmffile} &
\hspace{-0.8cm}
\begin{fmffile}{goldgzgz}
	        \begin{fmfgraph*}(60,50)
	            \fmfleft{i,j}
	            \fmfright{k,l}
	              \fmf{dashes,label=$G_\zp$,l.s=left,l.d=2}{i,v1}
		    \fmf{dashes,label=$G_\zp$,l.s=right,l.d=2}{j,v1}
		    \fmf{dashes,label=$G_{Z/W^+}$,l.s=right,l.d=2}{k,v1}
		    \fmf{dashes,label=$G_{Z/W^-}$,l.d=2}{v1,l}
	        \end{fmfgraph*}
\end{fmffile} &
\hspace{0.4cm}
\begin{fmffile}{goldhz}
	        \begin{fmfgraph*}(60,50)
	            \fmfleft{i,j}
	            \fmfright{k}
	            \fmf{dashes,label=$G_{Z/W^+}$,l.s=left,l.d=2}{i,v1}
		    \fmf{dashes,label=$G_{Z/W^-}$,l.s=right,l.d=2}{j,v1}
		    \fmf{dashes,label=$h_i$,l.d=2}{k,v1}
	        \end{fmfgraph*}
\end{fmffile}
\\
$i2\gx^2 g_{\mu\nu}$ &
$-i3\gx^2\frac{R_{21}^2 m_1^2+R_{22}^2 m_2^2}{\mzp^2}$ &
$-ig\gx\frac{\sin(2\alpha)(m_1^2-m_2^2)}{2M M_W}$ & 
$-ig\frac{m_i^2}{2M_W^2}R_{1i}$\\\hline
\end{tabular}
\end{center}
\caption{The vertices relevant for the calculation of amplitudes in the $R_\xi$ gauge within the vector dark matter model. The Goldstone field of the U(1)$_X$ symmetry is denoted by $G_\zp$.}
\label{vertices}
\end{table}
\begin{table}[h]
\begin{center}
\tetab
\begin{tabular}{|c|c|c|c|}
\hline
\vspace{-0.5cm}
& & & 
\\
\begin{fmffile}{bfgverhexgx}
	        \begin{fmfgraph*}(60,50)
	            \fmfleft{i,j}
	            \fmfright{k}
	            \fmf{dashes,label=$G_\zp$,l.d=7,l.s=left}{v1,i}
		    \fmf{dashes,label=$\pthat{h}_i$,l.d=7}{j,v1}
		    \fmf{boson,label=$\zp$}{k,v1}
		    \marrow{a}{down}{bot}{$k$}{i,v1}
		    \marrow{b}{up}{top}{$p$}{j,v1}
	        \end{fmfgraph*}
\end{fmffile}
&
 \begin{fmffile}{bfgverhgexx}
	        \begin{fmfgraph*}(60,50)
	            \fmfleft{i,j}
	            \fmfright{k}
	            \fmf{dashes,label=$\pthat{G}_\zp$,l.d=7,l.s=left}{v1,i}
		    \fmf{dashes,label=$h_i$,label.d=7}{j,v1}
		    \fmf{boson,label=$\zp$}{k,v1}
		    \marrow{a}{down}{bot}{$k$}{i,v1}
		    \marrow{b}{up}{top}{$p$}{j,v1}
	        \end{fmfgraph*}
\end{fmffile} & 
 \begin{fmffile}{bfgverhexgg}
	        \begin{fmfgraph*}(60,50)
	            \fmfleft{i,j}
	            \fmfright{k}
	            \fmf{dashes,label=$G_\zp$,l.d=2}{i,v1}
		    \fmf{dashes,label=$G_\zp$,l.s=left,l.d=2}{j,v1}
		    \fmf{dashes,label=$\pthat{h}_i$}{k,v1}
	        \end{fmfgraph*}
\end{fmffile} &
   \begin{fmffile}{bfgverhgexg}
	        \begin{fmfgraph*}(60,50)
	            \fmfleft{i,j}
	            \fmfright{k}
	            \fmf{dashes,label=$\pthat{G}_\zp$,l.d=2}{i,v1}
		    \fmf{dashes,label=$G_\zp$,l.s=left,l.d=2}{j,v1}
		    \fmf{dashes,label=$h_i$}{k,v1}
	        \end{fmfgraph*}
\end{fmffile}
\\
$2\gx R_{2i}p_\mu$
& $-2\gx R_{2i} k_\mu$
& $-i\gx R_{2i}(\frac{m_i^2}{\mzp}+2\xi_Q\mzp)$
& $-i\gx R_{2i}(\frac{m_i^2}{\mzp}-\xi_Q\mzp)$
\\\hline
& &    & \vspace{-0.5cm} 
\\
   \begin{fmffile}{bfgvercchex}
	  \begin{fmfgraph*}(60,50)
		      \fmfleft{i,j}
		      \fmfright{k}
		      \fmf{dashes_arrow,label=$\bar{c}$}{v,i}
		      \fmf{dashes_arrow,label=$c$,l.s=right}{j,v}
		      \fmf{dashes,label=$\pthat{h}_i$}{v,k}
		  \end{fmfgraph*}
  \end{fmffile}
 &
  \begin{fmffile}{bfgverccgexgex}
		  \begin{fmfgraph*}(60,50)
		      \fmfleft{i,j}
		      \fmfright{k,l}
		      \fmf{dashes_arrow,label=$\bar{c}$}{v,i}
		      \fmf{dashes_arrow,label=$c$,l.s=right}{j,v}
		      \fmf{dashes,label=$\pthat{G}_\zp$,l.d=0.1}{k,v}
		      \fmf{dashes,label=$\pthat{G}_\zp$,l.s=left,l.d=0.1}{l,v}
		  \end{fmfgraph*}
  \end{fmffile}
  &
  \begin{fmffile}{bfgverhexgzgz}
	        \begin{fmfgraph*}(60,50)
	            \fmfleft{i,j}
	            \fmfright{k}
	            \fmf{dashes,label=$G_Z$,l.d=2}{i,v1}
		    \fmf{dashes,label=$G_Z$,l.s=left,l.d=2}{j,v1}
		    \fmf{dashes,label=$\pthat{h}_i$}{k,v1}
	        \end{fmfgraph*}
\end{fmffile}  &
  \begin{fmffile}{bfgverhexgwgw}
	        \begin{fmfgraph*}(60,50)
	            \fmfleft{i,j}
	            \fmfright{k}
	            \fmf{dashes,label=$G_W^+$,l.d=2}{i,v1}
		    \fmf{dashes,label=$G_W^-$,l.s=left,l.d=2}{j,v1}
		    \fmf{dashes,label=$\pthat{h}_i$}{k,v1}
	        \end{fmfgraph*}
\end{fmffile} 
 \\
$-i2\gx\mzp\xi_Q R_{2i}$ & 
$-i2\gx^2\xi_Q$ &
$-i\text{g} \left(\frac{m_i^2}{2M_W}+\frac{M_Z^2}{M_W}\xi\right)$ &
$-i\text{g} \left(\frac{m_i^2}{2M_W}+M_W\xi\right)$
\\\hline
\end{tabular}
\end{center}
\caption{Vertices in the covariant background field gauge that differ
  from their counterparts in the $R_\xi$ gauge. We follow the
  conventions of~\cite{Denner:1994nn}. A hat over a symbol denotes a background~field.}
\end{table}

\break

\section{Analytic Expressions of the PT Vertices}
\label{ptvertices}

In order to compute the PT vertices $\pthat V_{\mu\nu}^{h_i XX}$,
$\pthat V_\mu^{h_i XG_X}$ and $\pthat V^{h_i G_X G_X}$, we will
exploit the equivalence between the PT and the covariant background field
gauge for $\xi_Q=1$, as shown in figs.~\ref{fig:PThiXX},~\ref{fig:PThiXGX}
and~\ref{fig:PThiGXGX}. For our purposes, we only need to take into
account the absorptive parts. Note that these are absent for the
self-energies $\pthat \Pi^{G_X G_X}$ and $\pthat \Pi^{X G_X}$.  When
calculating $\pthat V^{h_i \zp\zp}_{\mu\nu}$ and
$\pthat V_\mu^{h_i \zp G_X}$, we omit all those terms that vanish upon
contraction with the polarization vectors of the external $X$
bosons. To this end, the following abbreviations will be used:
\begin{equation}
\beta_X=\sqrt{1-4M_X^2/s}\ ,\;\;\beta_{h_{W/Z}}=\sqrt{1-4m_{W/Z}^2/s}\
,\;\;\lambda(x,y,z)=(x-y-z)^2-4yz \,. 
\end{equation}

\bigskip
\begin{center}
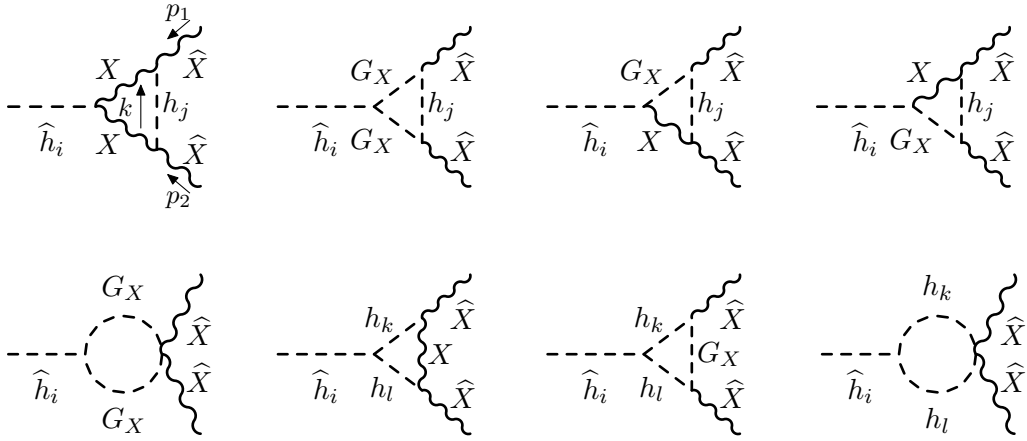
\begin{figure}[htb]
\setlength{\tabcolsep}{5pt}
\renewcommand{\arraystretch}{1.1}
\begin{tabular}{cccc}
\begin{fmffile}{ver1}
	          \begin{fmfgraph*}(80,60)
	          	   \fmfleft{v1}
	             \fmfright{i,j}
		    \fmf{dashes,label=$\pthat{h}_i$,tension=1.4}{v1,v2}
		    \fmf{boson,label=$X$,label.side=right,l.d=3}{v2,vu}
		    \marrow{a}{left}{lft}{$k$}{vu,vd}
		    \fmf{boson,label=$X$,label.side=left,l.d=3}{v2,vd}
		    \fmf{dashes,label=$h_j$,tension=0.1,label.side=right,l.d=2}{vu,vd}
		    \fmf{boson,label=$\pthat\zp$,tension=1.4,label.side=left,l.d=2}{vu,i}
		    \marrow{b}{up}{top}{$p_1$}{j,vd}
		    \marrow{c}{down}{bot}{$p_2$}{i,vu}
		    \fmf{boson,label=$\pthat\zp$,tension=1.4,label.side=right,l.d=2}{vd,j}
	        \end{fmfgraph*}
\end{fmffile}  &
\begin{fmffile}{ver2}
	          \begin{fmfgraph*}(80,60)
	          	   \fmfleft{v1}
	             \fmfright{i,j}
		    \fmf{dashes,label=$\pthat{h}_i$}{v1,v2}
		    \fmf{dashes,label=$G_X$,label.side=right,l.d=2}{v2,vu}
		    \fmf{dashes,label=$G_X$,label.side=left,l.d=2}{v2,vd}
		    \fmf{dashes,label=$h_j$,tension=0.1,label.side=right,l.d=2}{vu,vd}
		    \fmf{boson,label=$\pthat\zp$,tension=1,label.side=left,l.d=2}{vu,i}
		    \fmf{boson,label=$\pthat\zp$,tension=1,label.side=right,l.d=2}{vd,j}
	        \end{fmfgraph*}
\end{fmffile}  &
\begin{fmffile}{ver3}
	          \begin{fmfgraph*}(80,60)
	          	   \fmfleft{v1}
	             \fmfright{i,j}
		    \fmf{dashes,label=$\pthat{h}_i$}{v1,v2}
		    \fmf{boson,label=$X$,label.side=right,l.d=3}{v2,vu}
		    \fmf{dashes,label=$G_X$,label.side=left,l.d=2}{v2,vd}
		    \fmf{dashes,label=$h_j$,tension=0.1,label.side=right,l.d=2}{vu,vd}
		    \fmf{boson,label=$\pthat\zp$,tension=1,label.side=left,l.d=2}{vu,i}
		    \fmf{boson,label=$\pthat\zp$,tension=1,label.side=right,l.d=2}{vd,j}
	        \end{fmfgraph*}
\end{fmffile}  &
\begin{fmffile}{ver4}
	          \begin{fmfgraph*}(80,60)
	          	   \fmfleft{v1}
	             \fmfright{i,j}
		    \fmf{dashes,label=$\pthat{h}_i$}{v1,v2}
		    \fmf{dashes,label=$G_X$,label.side=right,l.d=2}{v2,vu}
		    \fmf{boson,label=$\zp$,label.side=left,l.d=3}{v2,vd}
		    \fmf{dashes,label=$h_j$,tension=0.1,label.side=right,l.d=2}{vu,vd}
		    \fmf{boson,label=$\pthat\zp$,tension=1,label.side=left,l.d=2}{vu,i}
		    \fmf{boson,label=$\pthat\zp$,tension=1,label.side=right,l.d=2}{vd,j}
	        \end{fmfgraph*}
\end{fmffile}  
\vspace{1cm}\\
\begin{fmffile}{ver5}
	          \begin{fmfgraph*}(80,60)
	          	   \fmfleft{v1}
	             \fmfright{i,j}
		    \fmf{dashes,label=$\pthat{h}_i$}{v1,vl}
		    \fmf{dashes,left,label=$G_X$,tension=0.5}{vl,vr,vl}
		    \fmf{boson,label=$\pthat\zp$,l.d=2}{vr,i}
		    \fmf{boson,label=$\pthat\zp$,l.d=2}{vr,j}
	        \end{fmfgraph*}
\end{fmffile} & \begin{fmffile}{ver6}
	          \begin{fmfgraph*}(80,60)
	          	   \fmfleft{v1}
	             \fmfright{i,j}
		    \fmf{dashes,label=$\pthat{h}_i$}{v1,v2}
		    \fmf{dashes,label=$h_l$,label.side=right,l.d=2}{v2,vu}
		    \fmf{dashes,label=$h_k$,label.side=left,l.d=2}{v2,vd}
		    \fmf{boson,label=$\zp$,tension=0.1,label.side=right,l.d=2}{vu,vd}
		    \fmf{boson,label=$\pthat\zp$,tension=1,label.side=left,l.d=2}{vu,i}
		    \fmf{boson,label=$\pthat\zp$,tension=1,label.side=right,l.d=2}{vd,j}
	        \end{fmfgraph*}
\end{fmffile} & \begin{fmffile}{ver7}
	          \begin{fmfgraph*}(80,60)
	          	   \fmfleft{v1}
	             \fmfright{i,j}
		    \fmf{dashes,label=$\pthat{h}_i$}{v1,v2}
		    \fmf{dashes,label=$h_l$,label.side=right,l.d=2}{v2,vu}
		    \fmf{dashes,label=$h_k$,label.side=left,l.d=2}{v2,vd}
		    \fmf{dashes,label=$G_X$,tension=0.1,label.side=right,l.d=2}{vu,vd}
		    \fmf{boson,label=$\pthat\zp$,tension=1,label.side=left,l.d=2}{vu,i}
		    \fmf{boson,label=$\pthat\zp$,tension=1,label.side=right,l.d=2}{vd,j}
	        \end{fmfgraph*}
\end{fmffile}  & \begin{fmffile}{ver8}
	          \begin{fmfgraph*}(80,60)
	          	   \fmfleft{v1}
	             \fmfright{i,j}
		    \fmf{dashes,label=$\pthat{h}_i$}{v1,vl}
		    \fmf{dashes,left,label=$h_k$,tension=0.5}{vl,vr}
		    \fmf{dashes,left,label=$h_l$,tension=0.5}{vr,vl}
		    \fmf{boson,label=$\pthat\zp$,l.d=2}{vr,i}
		    \fmf{boson,label=$\pthat\zp$,l.d=2}{vr,j}
	        \end{fmfgraph*}
\end{fmffile} 
\end{tabular}
\caption{Feynman diagrams related to $\pthat
  V_{\mu\nu}^{h_iXX}$.\label{fig:PThiXX}}
\end{figure}
\end{center}

\begin{center}
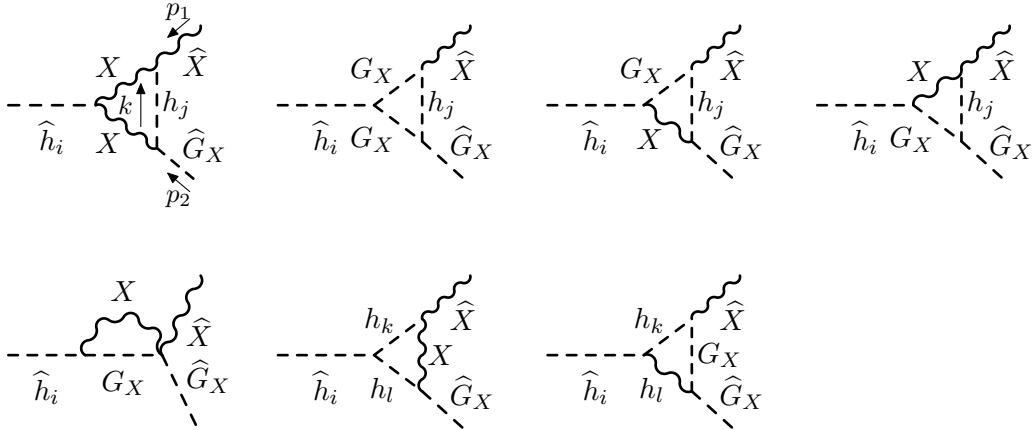
\begin{figure}[htb]
\setlength{\tabcolsep}{5pt} 
\renewcommand{\arraystretch}{0.3}
\begin{tabular}{cccc}        
\begin{fmffile}{xver1}
	          \begin{fmfgraph*}(80,60)
	          	   \fmfleft{v1}
	             \fmfright{i,j}
		    \fmf{dashes,label=$\pthat h_i$,tension=1.4}{v1,v2}
		    \fmf{boson,label=$X$,label.side=right,l.d=3}{v2,vu}
		    \marrow{a}{left}{lft}{$k$}{vu,vd}
		    \fmf{boson,label=$X$,label.side=left,l.d=3}{v2,vd}
		    \fmf{dashes,label=$h_j$,tension=0.1,l.d=2}{vu,vd}
		    \fmf{dashes,label=$\pthat G_X$,tension=1.4,label.side=left,l.d=2}{vu,i}
		    \marrow{b}{up}{top}{$p_1$}{j,vd}
		    \marrow{c}{down}{bot}{$p_2$}{i,vu}
		    \fmf{boson,label=$\pthat X$,tension=1.4,label.side=right,l.d=2}{vd,j}
	        \end{fmfgraph*}
\end{fmffile}  &
\begin{fmffile}{xver2}
	          \begin{fmfgraph*}(80,60)
	          	   \fmfleft{v1}
	             \fmfright{i,j}
		    \fmf{dashes,label=$\pthat h_i$}{v1,v2}
		    \fmf{dashes,label=$G_X$,label.side=right,l.d=2}{v2,vu}
		    \fmf{dashes,label=$G_X$,label.side=left,l.d=2}{v2,vd}
		    \fmf{dashes,label=$h_j$,tension=0.1,label.side=right,l.d=2}{vu,vd}
		    \fmf{dashes,label=$\pthat G_X$,tension=1,label.side=left,l.d=2}{vu,i}
		    \fmf{boson,label=$\pthat X$,tension=1,label.side=right,l.d=2}{vd,j}
	        \end{fmfgraph*}
\end{fmffile}  &
\begin{fmffile}{xver3}
	          \begin{fmfgraph*}(80,60)
	          	   \fmfleft{v1}
	             \fmfright{i,j}
		    \fmf{dashes,label=$\pthat h_i$}{v1,v2}
		    \fmf{boson,label=$X$,label.side=right,l.d=3}{v2,vu}
		    \fmf{dashes,label=$G_X$,label.side=left,l.d=2}{v2,vd}
		    \fmf{dashes,label=$h_j$,tension=0.1,label.side=right,l.d=2}{vu,vd}
		    \fmf{dashes,label=$\pthat G_X$,tension=1,label.side=left,l.d=2}{vu,i}
		    \fmf{boson,label=$\pthat X$,tension=1,label.side=right,l.d=2}{vd,j}
	        \end{fmfgraph*}
\end{fmffile}  &
\begin{fmffile}{xver4}
	          \begin{fmfgraph*}(80,60)
	          	   \fmfleft{v1}
	             \fmfright{i,j}
		    \fmf{dashes,label=$\pthat h_i$}{v1,v2}
		    \fmf{dashes,label=$G_X$,label.side=right,l.d=2}{v2,vu}
		    \fmf{boson,label=$\zp$,label.side=left,l.d=3}{v2,vd}
		    \fmf{dashes,label=$h_j$,tension=0.1,label.side=right,l.d=2}{vu,vd}
		    \fmf{dashes,label=$\pthat G_X$,tension=1,label.side=left,l.d=2}{vu,i}
		    \fmf{boson,label=$\pthat X$,tension=1,label.side=right,l.d=2}{vd,j}
	        \end{fmfgraph*}
\end{fmffile}  
\vspace{1cm}\\\\
\begin{fmffile}{xver5}
	          \begin{fmfgraph*}(80,60)
	          	   \fmfleft{v1}
	             \fmfright{i,j}
		    \fmf{dashes,label=$\pthat h_i$}{v1,vl}
		    \fmf{boson,left,label=$X$,tension=0.5}{vl,vr}
		    \fmf{dashes,label=$G_X$,tension=0.5}{vl,vr}
		    \fmf{dashes,label=$\pthat G_X$,l.d=2}{vr,i}
		    \fmf{boson,label=$\pthat X$,l.d=2}{vr,j}
	        \end{fmfgraph*}
\end{fmffile} & \begin{fmffile}{xver6}
	          \begin{fmfgraph*}(80,60)
	          	   \fmfleft{v1}
	             \fmfright{i,j}
		    \fmf{dashes,label=$\pthat h_i$}{v1,v2}
		    \fmf{dashes,label=$h_l$,label.side=right,l.d=2}{v2,vu}
		    \fmf{dashes,label=$h_k$,label.side=left,l.d=2}{v2,vd}
		    \fmf{boson,label=$\zp$,tension=0.1,label.side=right,l.d=2}{vu,vd}
		    \fmf{dashes,label=$\pthat G_X$,tension=1,label.side=left,l.d=2}{vu,i}
		    \fmf{boson,label=$\pthat X$,tension=1,label.side=right,l.d=2}{vd,j}
	        \end{fmfgraph*}
\end{fmffile}  &
\begin{fmffile}{xver7}
	          \begin{fmfgraph*}(80,60)
	          	   \fmfleft{v1}
	             \fmfright{i,j}
		    \fmf{dashes,label=$\pthat h_i$}{v1,v2}
		    \fmf{boson,label=$h_l$,label.side=right,l.d=2}{v2,vu}
		    \fmf{dashes,label=$h_k$,label.side=left,l.d=2}{v2,vd}
		    \fmf{dashes,label=$G_X$,tension=0.1,label.side=right,l.d=2}{vu,vd}
		    \fmf{dashes,label=$\pthat G_X$,tension=1,label.side=left,l.d=2}{vu,i}
		    \fmf{boson,label=$\pthat X$,tension=1,label.side=right,l.d=2}{vd,j}
	        \end{fmfgraph*}
\end{fmffile}  &
\end{tabular}
\caption{Feynman diagrams contributing to $\pthat
  V_\mu^{h_iXG_X}$.\label{fig:PThiXGX} }
\end{figure}
\end{center}

${}$\vspace{2cm}${}$

\begin{center}	
\begin{figure}[htb]
\setlength{\tabcolsep}{5pt}
\renewcommand{\arraystretch}{0.5}
\begin{tabular}{cccc}        
\begin{fmffile}{gver1}
	          \begin{fmfgraph*}(80,60)
	          	   \fmfleft{v1}
	             \fmfright{i,j}
		    \fmf{dashes,label=$\pthat{h}_i$,tension=1.4}{v1,v2}
		    \fmf{boson,label=$X$,label.side=right,l.d=3}{v2,vu}
		    \marrow{a}{left}{lft}{$k$}{vu,vd}
		    \fmf{boson,label=$X$,label.side=left,l.d=3}{v2,vd}
		    \fmf{dashes,label=$h_j$,tension=0.1,l.d=2}{vu,vd}
		    \fmf{dashes,label=$\pthat{G}_X$,tension=1.4,label.side=left,l.d=2}{vu,i}
		    \marrow{b}{up}{top}{$p_1$}{j,vd}
		    \marrow{c}{down}{bot}{$p_2$}{i,vu}
		    \fmf{dashes,label=$\pthat{G}_X$,tension=1.4,label.side=right,l.d=2}{vd,j}
	        \end{fmfgraph*}
\end{fmffile}  &
\begin{fmffile}{gver2}
	          \begin{fmfgraph*}(80,60)
	          	   \fmfleft{v1}
	             \fmfright{i,j}
		    \fmf{dashes,label=$\pthat{h}_i$}{v1,v2}
		    \fmf{dashes,label=$G_X$,label.side=right,l.d=2}{v2,vu}
		    \fmf{dashes,label=$G_X$,label.side=left,l.d=2}{v2,vd}
		    \fmf{dashes,label=$h_j$,tension=0.03,label.side=right,l.d=2}{vu,vd}
		    \fmf{dashes,label=$\pthat{G}_X$,tension=1,label.side=left,l.d=2}{vu,i}
		    \fmf{dashes,label=$\pthat{G}_X$,tension=1,label.side=right,l.d=2}{vd,j}
	        \end{fmfgraph*}
\end{fmffile}  &
\begin{fmffile}{gver3}
	          \begin{fmfgraph*}(80,60)
	          	   \fmfleft{v1}
	             \fmfright{i,j}
		    \fmf{dashes,label=$\pthat{h}_i$}{v1,v2}
		    \fmf{boson,label=$X$,label.side=right,l.d=3}{v2,vu}
		    \fmf{dashes,label=$G_X$,label.side=left,l.d=2}{v2,vd}
		    \fmf{dashes,label=$h_j$,tension=0.1,label.side=right,l.d=2}{vu,vd}
		    \fmf{dashes,label=$\pthat{G}_X$,tension=1,label.side=left,l.d=2}{vu,i}
		    \fmf{dashes,label=$\pthat{G}_X$,tension=1,label.side=right,l.d=2}{vd,j}
	        \end{fmfgraph*}
\end{fmffile}  &
\begin{fmffile}{gver4}
	          \begin{fmfgraph*}(80,60)
	          	   \fmfleft{v1}
	             \fmfright{i,j}
		    \fmf{dashes,label=$\pthat{h}_i$}{v1,v2}
		    \fmf{dashes,label=$G_X$,label.side=right,l.d=2}{v2,vu}
		    \fmf{boson,label=$\zp$,label.side=left,l.d=3}{v2,vd}
		    \fmf{dashes,label=$h_j$,tension=0.1,label.side=right,l.d=2}{vu,vd}
		    \fmf{dashes,label=$\pthat{G}_X$,tension=1,label.side=left,l.d=2}{vu,i}
		    \fmf{dashes,label=$\pthat{G}_X$,tension=1,label.side=right,l.d=2}{vd,j}
	        \end{fmfgraph*}
\end{fmffile}  
\vspace{1cm}\\\\
\begin{fmffile}{gver5}
	          \begin{fmfgraph*}(80,60)
	          	   \fmfleft{v1}
	             \fmfright{i,j}
		    \fmf{dashes,label=$\pthat{h}_i$}{v1,vl}
		    \fmf{dashes,left,label=$G_X$,tension=0.5}{vl,vr,vl}
		    \fmf{dashes,label=$\pthat{G}_X$,l.d=2}{vr,i}
		    \fmf{dashes,label=$\pthat{G}_X$,l.d=2}{vr,j}
	        \end{fmfgraph*}
\end{fmffile}  &
\begin{fmffile}{gver6}
	          \begin{fmfgraph*}(80,60)
	          	   \fmfleft{v1}
	             \fmfright{i,j}
		    \fmf{dashes,label=$\pthat{h}_i$}{v1,vl}
		    \fmf{boson,left,label=$\zp$,tension=0.5}{vl,vr,vl}
		    \fmf{dashes,label=$\pthat{G}_X$,l.d=2}{vr,i}
		    \fmf{dashes,label=$\pthat{G}_X$,l.d=2}{vr,j}
	        \end{fmfgraph*} 
\end{fmffile} & \begin{fmffile}{gver7}
	          \begin{fmfgraph*}(80,60)
	          	   \fmfleft{v1}
	             \fmfright{i,j}
		    \fmf{dashes,label=$\pthat{h}_i$}{v1,vl}
		    \fmf{dashes_arrow,left,label=$\bar c$,tension=0.5}{vl,vr}
		    \fmf{dashes_arrow,left,label=$c$,tension=0.5}{vr,vl}
		    \fmf{dashes,label=$\pthat{G}_X$,l.d=2}{vr,i}
		    \fmf{dashes,label=$\pthat{G}_X$,l.d=2}{vr,j}
	        \end{fmfgraph*}
\end{fmffile} & 
\begin{fmffile}{gver8}
	          \begin{fmfgraph*}(80,60)
	          	   \fmfleft{v1}
	             \fmfright{i,j}
		    \fmf{dashes,label=$\pthat{h}_i$}{v1,v2}
		    \fmf{dashes,label=$h_l$,label.side=right,l.d=2}{v2,vu}
		    \fmf{dashes,label=$h_k$,label.side=left,l.d=2}{v2,vd}
		    \fmf{dashes,label=$\zp$,tension=0.03,label.side=right,l.d=2}{vu,vd}
		    \fmf{dashes,label=$\pthat{G}_X$,tension=1,label.side=left,l.d=2}{vu,i}
		    \fmf{dashes,label=$\pthat{G}_X$,tension=1,label.side=right,l.d=2}{vd,j}
	        \end{fmfgraph*}
\end{fmffile} 
\vspace{1cm}\\\\
\begin{fmffile}{gver9}
	          \begin{fmfgraph*}(80,60)
	          	   \fmfleft{v1}
	             \fmfright{i,j}
		    \fmf{dashes,label=$\pthat{h}_i$}{v1,v2}
		    \fmf{dashes,label=$h_l$,label.side=right,l.d=2}{v2,vu}
		    \fmf{dashes,label=$h_k$,label.side=left,l.d=2}{v2,vd}
		    \fmf{dashes,label=$G_X$,tension=0.03,label.side=right,l.d=2}{vu,vd}
		    \fmf{dashes,label=$\pthat{G}_X$,tension=1,label.side=left,l.d=2}{vu,i}
		    \fmf{dashes,label=$\pthat{G}_X$,tension=1,label.side=right,l.d=2}{vd,j}
	        \end{fmfgraph*}
\end{fmffile} &
\begin{fmffile}{gver10}
	          \begin{fmfgraph*}(80,60)
	          	   \fmfleft{v1}
	             \fmfright{i,j}
		    \fmf{dashes,label=$\pthat{h}_i$}{v1,vl}
		    \fmf{dashes,left,label=$h_k$,tension=0.5}{vl,vr}
		    \fmf{dashes,left,label=$h_l$,tension=0.5}{vr,vl}
		    \fmf{dashes,label=$\pthat{G}_X$,l.d=2}{vr,i}
		    \fmf{dashes,label=$\pthat{G}_X$,l.d=2}{vr,j}
	        \end{fmfgraph*}
\end{fmffile}  &
\begin{fmffile}{gver11}
	          \begin{fmfgraph*}(80,60)
	          	   \fmfleft{v1}
	             \fmfright{i,j}
		    \fmf{dashes,label=$\pthat{h}_i$}{v1,vl}
		    \fmf{dashes,left,label=$G_Z$,tension=0.5}{vl,vr,vl}
		    \fmf{dashes,label=$\pthat{G}_X$,l.d=2}{vr,i}
		    \fmf{dashes,label=$\pthat{G}_X$,l.d=2}{vr,j}
	        \end{fmfgraph*}
\end{fmffile} &
\begin{fmffile}{gver12}
	          \begin{fmfgraph*}(80,60)
	          	   \fmfleft{v1}
	             \fmfright{i,j}
		    \fmf{dashes,label=$\pthat{h}_i$}{v1,vl}
		    \fmf{dashes,left,label=$G_W^+$,tension=0.5}{vl,vr}
		    \fmf{dashes,left,label=$G_W^-$,tension=0.5}{vr,vl}
		    \fmf{dashes,label=$\pthat{G}_X$,l.d=1}{vr,i}
		    \fmf{dashes,label=$\pthat{G}_X$,l.d=1}{vr,j}
	        \end{fmfgraph*}
\end{fmffile}  
\end{tabular} 	
\caption{Feynman diagrams pertinent to $\pthat
  V^{h_iG_XG_X}$.\label{fig:PThiGXGX}}
\end{figure}
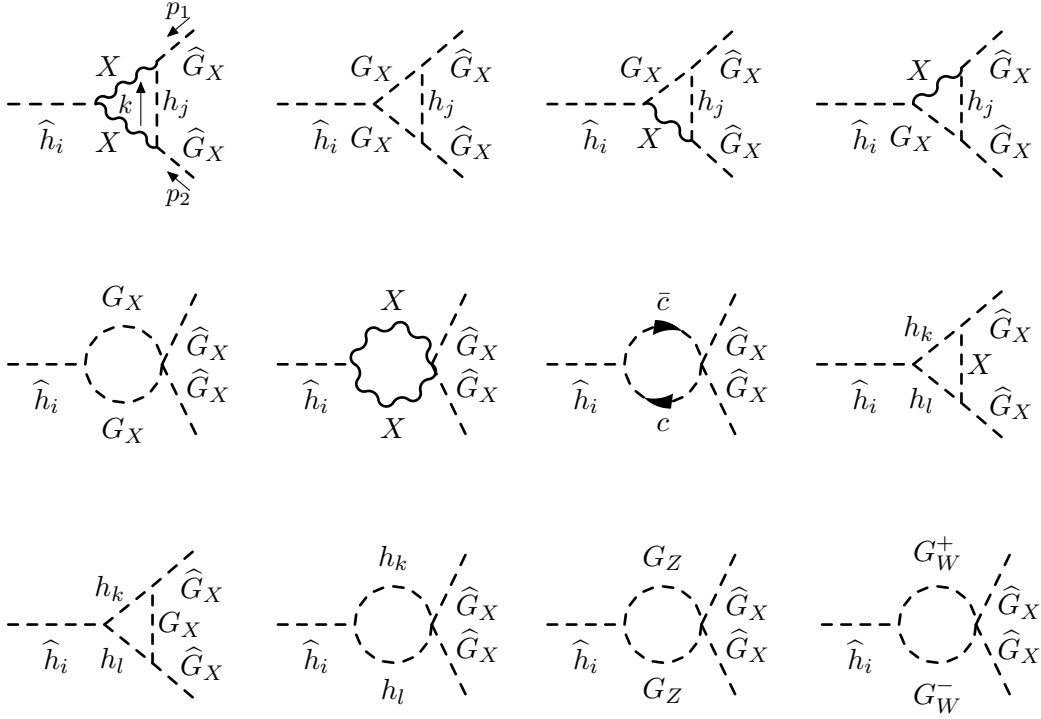	
\end{center}

After a lengthy but straightforward computation of the Feynman
diagrams shown in figs.~\ref{fig:PThiXX},~\ref{fig:PThiXGX}
and~\ref{fig:PThiGXGX}, the following analytic results for the absorptive parts
of the PT vertices $\pthat V_{\mu\nu}^{h_i XX}$,
$\pthat V_\mu^{h_i XG_X}$ and $\pthat V^{h_i G_X G_X}$ are obtained:
\begin{align*}
&\pthat V^{h_i\zp\zp}_{\mu\nu} = \sum_j\frac{\gx^3 R_{2i}R_{2j}^2}{8\pi s^{3}\beta_X^3}\theta(s-4\mzp^2)\bigg[
-s\beta_X^2m_j^2(m_i^2+2M^2)g_{\mu\nu}
-2\big[8\mzp^6\\ghostghost
&-2\mzp^4(9s-2m_i^2+2m_j^2)+\mzp^2\big(4s(s+m_j^2)-m_i^2(s+2m_j^2)\big)+2m_i^2m_j^2s\big]r_{2\mu}r_{1\nu}\\
&+\log\left[1+\frac{s\beta_X^2}{m_j^2}\right]
\Big(\big[16\mzp^6-4\mzp^4(2m_j^2+s)+2\mzp^2m_j^2(s+m_j^2-2m_i^2)\\ &+m_i^2m_j^2(s+m_j^2)\big]g_{\mu\nu} - 2 \frac{m_j^2}{s\beta_X^2}\big[16\mzp^6+4\mzp^4(2m_i^2-m_j^2-7s)\\
&+\mzp^2\big(2s(3s+2m_j^2)-2m_i^2(3s+m_j^2)\big)+sm_i^2(s+2m_j^2)\big]r_{2\mu}r_{1\nu}\bigg]\\
& +\sum_{k,\,l}\frac{i\gx^2 R_{2k}R_{2l}V^h_{ikl}}{16\pi s^{4}\beta_X^4}\theta[s-(m_k+m_l)^2]\bigg[\lambda^{1/2}(s,m_k^2,m_l^2)\\\times
&\Big(-s^2\beta^2_X(4\mzp^2-m_k^2-m_l^2)g_{\mu\nu}+4\big[\mzp^2(3s+m_k^2+m_l^2)-s(m_k^2+m_l^2)\big]r_{2\mu}r_{1\nu}\Big)\\
&+\frac{2}{s\beta_X}\log\left[\frac{s-(m_k^2+m_l^2)-s\beta_X\lambda^{1/2}(s,m_k^2,m_l^2)}{s-(m_k^2+m_l^2)+s\beta_X\lambda^{1/2}(s,m_k^2,m_l^2)}\right]\\\times
&\Big(s\beta_X^2\big[-8\mzp^4s+\mzp^2\big(2s(m_k^2+m_l^2)-(m_k^2-m_l^2)^2+s^2\big)-m_k^2m_l^2s\big]g_{\mu\nu}\tag{\stepcounter{equation}\theequation}\\
&+2\Big[2\mzp^4\big(2s(m_k^2+m_l^2)-3(m_k^2-m_l^2)^2+s^2\big)\\&+s\mzp^2\big(s^2-4s(m_k^2+m_l^2)+3(m_k^4+m_l^4)-8m_k^2m_l^2\big)+2m_k^2m_l^2s^2\Big]r_{2\mu}r_{1\nu}
\Big)
\bigg]\,,
\end{align*}
\begin{align*}
&\pthat V_\mu^{h_i XG_X} =
-\sum_j\frac{i\gx^3 R_{2i}R_{2j}^2 r_{2\mu}}{8\pi M_X^2 s^2\beta_X}\theta[s-4\mzp^2]
\bigg[10M_X^2+M_X^2(m_i^2-2m_j^2-4s)-m_i^2m_j^2\\
&+\beta_X^{-2}\big[8M_X^6-2M_X^4(s-5m_j^2)+M_X^2m_j^2(m_i^2-2m_j^2-4s)-m_i^2m_j^4\big]\log\hspace{-0.05cm}\bigg[1+\frac{s\beta_X^2}{m_j^2}\bigg]\bigg]\\
&+\sum_{k,\,l}\frac{\gx^2 R_{2k}R_{2l} V^h_{ikl}r_{2\mu}}{16\pi\beta_X^2 s^3\mzp^2}\theta[s-(m_k+m_l)^2]
\bigg[\lambda^{1/2}(s,m_k^2,m_l^2)s[2M_X^2-(m_k^2+m_l^2)]\\
& + \beta_X^{-2}\log\left[\frac{s-(m_k^2+m_l^2)-s\beta_X\lambda^{1/2}(s,m_k^2,m_l^2)}{s-(m_k^2+m_l^2)+s\beta_X\lambda^{1/2}(s,m_k^2,m_l^2)}\right]\tag{\stepcounter{equation}\theequation}\\\times
&\big[2m_k^2m_l^2s+\mzp^2\big(4s^2-3s(m_k^2+m_l^2)+2(m_k^2-m_l^2)^2\big)-12\mzp^4 s\big]\bigg]\,,
\end{align*}
\beq
\begin{split}
&\pthat V^{h_iG_XG_X} = 
\sum_j\frac{\gx^3 R_{2i}R_{2j}^2}{32\pi s\beta_X}\theta[s-4\mzp^2]\bigg[s\beta_X^2\big(8\mzp^4+3m_j^2(2\mzp^2+m_i^2)\big)\\
&-2\big[2\mzp^2m_j^2(-m_i^2+m_j^2+2s)+\mzp^4(m_i^2-4m_j^2)+m_i^2m_j^4-6\mzp^6\big]\times\\
&\log\bigg[1-\frac{s\beta_X^2}{m_j^2}\bigg]\bigg]
 +\sum_{k,\,l}\frac{\gx V^h_{ikl}}{128\pi s M_X^2}\theta[s-(m_k+m_l)^2]\bigg[\lambda^{1/2}(s,m_k^2,m_l^2)\big[2\gx R_{2k}R_{2l}\times\\
 &\big(4M^2+m_1^2+m_2^2+\cos(2\alpha)(m_2^2-m_1^2)\big)+g\frac{M_X}{M_W}R_{1k}R_{1l}(m_1^2-m_2^2)\sin(2\alpha)\big]\\
 &+8\beta_X^{-1/2}\log\left[\frac{s-(m_k^2+m_l^2)-s\beta_X\lambda^{1/2}(s,m_k^2,m_l^2)}{s-(m_k^2+m_l^2)+s\beta_X\lambda^{1/2}(s,m_k^2,m_l^2)}\right]\times\\
 &\big[3\mzp^4-\mzp^2(2s-m_k^2-m_l^2)-m_k^2m_l^2\big]\bigg]\\
 +&\sum_j\frac{g^2g_X R_{1i}R_{1j}R_{2j}m_j^2}{32\pi M_Z M_X}\theta(s-4M_Z^2)\beta_Z\\
 +&\sum_j\frac{g^2g_X R_{1i}(m_2^2-m_1^2)\sin(2\alpha)}{128\pi M_W^2 M_X}\theta[s-4M_W^2]\beta_W\,.
\end{split}
\eeq
\section{Generalized Equivalence Theorem for Resummed Amplitudes}
\label{getapp}

The proof of the Generalized Equivalence Theorem (GET) for the
amplitude $\mathcal{A}_{XX\rightarrow \bar ff}$ goes along the lines
of the one given in~\cite{Papavassiliou:1997pb} for the SM process
$\bar f f \rightarrow ZZ$. It is based on the classical WIs that are
satisfied by the PT Higgs self-energies and the PT $hZZ$ vertex.

In order to prove the GET for the process $XX\rightarrow \bar ff$,
\begin{equation}
\begin{split}
\mathcal{A}_{X_L(p_1)X_L(p_2)\rightarrow \bar f f}=
&-\mathcal{A}_{G_X(p_1)G_X(p_2)\rightarrow \bar f
  f}-i\mathcal{A}_{x^\mu(p_1)G_X(p_2)\rightarrow \bar f f}\\ 
&-i\mathcal{A}_{G_X(p_1)x^\nu(p_2)\rightarrow \bar f
  f}+\mathcal{A}_{x^\mu(p_1)x^\nu(p_2)\rightarrow \bar f f}\,, 
\end{split}
\end{equation}
we have to check that the following identity holds true (summation
over the repeated indices $i,j,k$ implied): 
\begin{equation}
   \label{eq:Depsilon}
\begin{split}
&\epsilon^\mu(p_1)\epsilon^\nu(p_2)(V_{\mu\nu}^{h_iXX}+\pthat
V_{\mu\nu}^{h_iXX})\pthat \Delta_{ij}V^{h_j\bar ff} =
-(V^{h_iG_XG_X}+\pthat V^{h_iG_XG_X})\pthat \Delta_{ij}V^{h_j\bar
  ff}\\ 
&-ix^\mu(p_1)(V_{\mu}^{h_iXG_X}+\pthat V_{\mu}^{h_iXG_X})\pthat
\Delta_{ij}V^{h_j\bar ff} -ix^\nu(p_2)(V_{\nu}^{h_iG_XX}+\pthat
V_{\nu}^{h_iG_XX})\pthat \Delta_{ij}V^{h_j\bar ff}\\ 
&+ x^\mu(p_1)x^\nu(p_2)(V_{\mu\nu}^{h_iXX}+\pthat 
V_{\mu\nu}^{h_iXX})\pthat \Delta_{ij}V^{h_j\bar ff}\;. 
\end{split}
\end{equation} 
Writing $\epsilon^\mu(p_{1,2})=p^\mu_{1,2}/M_X+x^\mu(p_{1,2})$
and using (\ref{WI2}) and (\ref{WI3}), one arrives from~\eqref{eq:Depsilon} at an 
equivalent expression
{\small\begin{equation}
  \label{eq:Did}
\begin{split}
&i\frac{g}{M_X}\bigg(R_{2k}\Big[(s-m_k^2)\delta_{ki}+\pthat\Pi_{ki}\Big]+
R_{2i}\Big[-p_1^2+\pthat\Pi^{G_X
  G_X}(p_1^2)-p_2^2+\pthat\Pi^{G_X
  G_X}(p_2^2)\Big]\bigg)\pthat\Delta_{ij}V^{h_j\bar ff}\\ 
&-\frac{g}{M_X}R_{2i}\bigg(x^\mu(p_1)\Big[iM_Xp_{1\mu}+\pthat\Pi_\mu^{X
  G_X}(p_1^2)\Big]+x^\nu(p_2)\Big[iM_Xp_{2\nu}+\pthat\Pi_\nu^{G_X
  X}(p_1^2)\Big]\bigg)\pthat\Delta_{ij}V^{h_j\bar ff}=0\,. 
\end{split}
\end{equation}}
\hspace{-2.5mm}
This last expression can be shown to be valid for both the tree-level
and its PT extension.  To do so,  we first observe that
$(s-m_k^2)\delta_{ki}+\pthat\Pi_{ki}=\pthat\Delta_{ki}^{-1}$. Then,
using (\ref{xident}) and (\ref{WI4}), one finds that the main bulk
of the terms in~\eqref{eq:Did}
cancel. Thus, one is only left to prove that 
\beq
i\frac{g}{M_X}R_{2k}\delta_{kj}V^{h_j\bar ff}\ =\ 0\;. \eeq 
The latter is true due to the orthogonality of the mixing matrix~$R$.
This concludes our proof of the GET for the
amplitude $\mathcal{A}_{XX\rightarrow \bar ff}$.

\bibliography{biblio2}
\bibliographystyle{JHEP}

\end{document}